\documentclass[12pt]{iopart}
\pdfoutput=1
\usepackage{graphicx}
\usepackage{iopams}
 
\usepackage{color}
\usepackage[numbers,square, comma, sort&compress]{natbib}

\pdfoutput=1

\newcommand\eq[1]{Eq.~(\ref{#1})}
\begin{document}

\title[Exact results for the temperature-field behavior of a mean-field model]
{Exact results for the temperature-field behavior of the Ginzburg-Landau Ising type mean-field model}
\author{Daniel M Dantchev, Vassil M Vassilev,  \MakeLowercase{and} Peter A Djondjorov}
\address{Institute of Mechanics, Bulgarian Academy of Sciences\\
Acad. G. Bonchev St., Building 4, 1113 Sofia, Bulgaria}
\ead{daniel@imbm.bas.bg, vasilvas@imbm.bas.bg
and padjon@imbm.bas.bg}

\begin{abstract}
We investigate the dependence of the order parameter profile, local and total susceptibilities on both the temperature and external magnetic field within the mean-filed Ginzburg-Landau Ising type model. We study the case of a film geometry when the boundaries of the film exhibit strong adsorption to one of the phases (components) of the system. We do that using general scaling arguments and deriving exact analytical results for the corresponding scaling functions of these quantities.  In addition, we examine their behavior in the capillary condensation regime. Based on the derived exact analytical expressions we obtained an unexpected result -- the existence of a region in the phase transitions line where the system jumps below its bulk critical temperature from a less dense gas to a more dense gas before switching on continuously into the usual jump from gas to liquid state in the middle of the system.
It is also demonstrated that on the capillary condensation line one of the coexisting local susceptibility profiles is with one maximum, whereas the other one is with two local maxima centered, approximately, around the two gas-liquid interfaces in the system.

\vspace{2pc}
\end{abstract}


\maketitle

\vspace{2pc}
\noindent %\hspace{14pt} %
{\normalsize \textbf{Contents}}
%\tableofcontents

\contentsline {section}{\numberline {1}Introduction}{3}
\contentsline {section}{\numberline {2}The Ginzburg-Landau mean-field model and basic expressions defining the finite-size quantities of the system}{5}
\contentsline {section}{\numberline {3}Analytical results for the scaling behavior of the order parameter profiles}{7}
\contentsline {subsection}{\numberline {3.1}Analytical representation of the order parameter profiles in the case of zero field}{7}
\contentsline {subsection}{\numberline {3.2}Analytical results for the order parameter profiles in the case of nonzero field}{9}
\contentsline {section}{\numberline {4}Analytical results for the scaling behavior of the susceptibility}{13}
\contentsline {subsection}{\numberline {4.1}Exact results for the local susceptibility profiles in the case of nonzero field}{13}
\contentsline {subsection}{\numberline {4.2}Exact results for the local susceptibility profiles in the case of zero field}{15}
\contentsline {subsection}{\numberline {4.3}Exact results for the total susceptibility in the case of nonzero field}{16}
\contentsline {subsection}{\numberline {4.4}Exact results for the total susceptibility in the case of zero field}{18}
\contentsline {section}{\numberline {5}Discussion and concluding remarks}{18}
\contentsline {section}{\numberline {}Appendix}{19}
\contentsline {section}{\numberline {}References}{21}

\newpage

\section{Introduction}

The understanding of the phase behavior of fluids confined in narrow regions, including the fluid mediated interactions, is of crucial importance in the physics of fluids in porous media, for colloidal physics, for many applications and modern technologies such as lubrication, adhesion or friction, in micro and nano-fluidics as well as for the proper interpretation of surface force experiments \cite{GGRS99,E90,E90book,BDT2000,BGK2003,P2006,BH2006,Bi2009,O2010,BK2010,I2011}.

In the current article the order parameter profile and its response functions to an externally applied ordering field $h$, as well as  the free energy will be investigated as functions of both the temperature $T$ and $h$ for the three-dimensional continuum mean-field Ising model with a film geometry $\infty^2 \times L$. We will consider the full $(T,h)$ plane for the case when the bounding surfaces of the system strongly prefer the ordered phase of the system which can be a simple fluid, magnetic system close to their respective critical point, or a binary liquid mixture close to its demixing point.
This is a standard model within which one studies phenomena like critical adsorption \cite{PL83, E90,FD95,THD2008,BU2001,REUM86,OO2012,MCS98,DSD2007,DMBD2009,C77,G85,TD2010,DRB2007},  wetting or drying \cite{C77,NF82,G85,BME87,Di88,SOI91}, surface phenomena \cite{Bb83,D86}, capillary condensation \cite{BME87,E90,BLM2002,REUM86,OO2012,DSD2007,YOO2013}, localization-delocalization phase transition \cite{PE90,PE92,BLM2003}, finite-size behavior of thin films \cite{KO72,NF83,FN81,BLM2003,NAFI83,REUM86,Ba83,C88,Ped90,PE90,BDT2000}, the thermodynamic Casimir effect \cite{K97,SHD2003,GaD2006,DSD2007}, etc.  One normally derives the results for $h=0$ analytically \cite{K97,GaD2006,DRB2009} while the $h$-dependence is studied numerically either at the bulk critical point of the system $T=T_c$, or along some specific isotherms -- see, e.g., \cite{SHD2003,MB2005,DSD2007,DRB2007,DRB2009,PE92,LTHD2014}.

In the current study we will present analytical results for the $h$-dependence of the model and will provide results for the full $(T,h)$ dependences of the order parameter, local and total susceptibilities in the $(T,h)$ plane. In what follows we will mainly use  the magnetic terminology but when considering capillary condensation we will also use that one of a simple fluid system in order better to reflect the physics of the obtained exact  mathematical results.  

In order to be more specific let us remind some facts and definitions pertinent to the above mentioned problems. 

If a fluid, or a magnetic system possesses a bounding surface its phase behavior as a function of its temperature $T$, excess chemical potential $\Delta \mu\propto h$ and the material characteristics of the surface are essentially enriched -- near the surface  one can have, e.g., phenomena of wetting or drying \cite{Di88}. In the vicinity of the bulk critical temperature $T_c$ of the bulk system, one observes a diversity of surface phase transitions \cite{Bb83,D86} of different kind in which the surface orders before, together, or after ordering in the bulk of the system, which is known as normal (or extraordinary), surface-bulk and ordinary surface phase transitions. For a simple fluid or for binary liquid
mixtures the wall generically prefers one of the fluid phases or one of the components. In the vicinity of the bulk critical 
point the last leads to the phenomenon of critical adsorption \cite{E90,FD95,THD2008,BU2001,REUM86,OO2012,MCS98,DSD2007,DMBD2009,C77,G85,TD2010,DRB2007}. Obviously, the surface breaks the spatial symmetry of the bulk system. The penetration depth of the effects due to the existence of a  bounding surface in the body of the system is set by the correlation length $\xi$ of the order parameter; $\xi$ becomes large, and theoretically diverges, in the vicinity of the bulk critical point $(T_c,h=0)$: $\xi(T\to T_c^{+},h=0)\simeq \xi_0^{+}\tau^{-\nu}$, $\tau=(T-T_c)/T_c$, and $\xi(T=T_c,h\to 0)\simeq \xi_{0,h} |h/(k_B T_c)|^{-\nu/\Delta}$, where $\nu$ and $\Delta$ are the usual critical exponents and $\xi_0^{+}$ and $\xi_{0,h}$ are the corresponding nonuniversal amplitudes of the correlation length along the $\tau$ and $h$ axes. When at least one of the spacial extensions of the system is finite, as in the $\infty^2 \times L$ film geometry we consider in the current article, one terms the corresponding system a finite system. If in such a system  $\xi$ becomes comparable to  $L$, the thermodynamic functions describing its behavior depend on the ratio $L/\xi$ and take a scaling form given by the finite-size scaling theory \cite{Ba83,C88,Ped90,PE90,BDT2000}. One observes, {\it inter alia}, shift of the critical point of the system \cite{NF83,FN81,BLM2003,NAFI83,REUM86} from $T_c$ to $T_{c,L}$. When in the finite system there is a phase transition of its own $T_{c,L}$ is a true critical point. Below $T_c$, if the confining walls of the film  are of the same material, the capillary condensation occurs \cite{E90,BLM2002,REUM86,OO2012,DSD2007,YOO2013} where, e.g., the liquid vapor coexistence line shifts away from the bulk coexistence into the one-phase regime. This coexistence line of first order transitions ends at a point $T_{c,L}$, which is normally identified with the capillary condensation  point $T_{\rm cap}$ that, on its turn, is considered to correspond to the highest temperature at which the entire capillary fills with liquid. In the current article we will demonstrate  that $T_{\rm cap}=T_{c,L}$ is not always true. We will show that, within the studied model, $T_{\rm cap}$ differs from $T_{c,L}$ on a scale determined by $L^{-1/\nu}$.

As already stated above, near a confining wall the symmetry between the two phases of a simple fluid or between the two components of the binary liquid mixture is violated in that one of these phases or components is preferred by the boundary. Thus, the order parameter profile, the density or the composition, becomes a function of the perpendicular coordinate $z$. This can be  modeled by considering local surface fields ${h}_1$ and ${h}_2$ acting solely on the surfaces of the system.  When the system undergoes a phase transition in its bulk in the presence of such surface ordering fields one speaks about the "{\em normal}" transition \cite{BD94}. It has been shown that it is equivalent, as far as the leading critical behavior is concerned, to the "{\it extraordinary}" transition \cite{D86,BD94} which is achieved by enhancing the surface couplings stronger than the bulk
couplings.  In the remainder of this article we will use the surface field picture. It has been demonstrated \cite{D86} that when $h_1h_2\ne 0$ for the leading critical behavior of the system is sufficient to investigate the limits $h_1, h_2 \to \pm \infty$. Obviously, there are two principal sub-cases $h_1 =
h_2 \to +\infty$, and $h_1 = -h_2 \to +\infty$ corresponding to $h_1h_2>0$ and $h_1h_2<0$.  One usually refers to the former case as the $(+,+)$ boundary
conditions and to the latter case as the $(+,-)$ boundary conditions. In the current article we will be only dealing with the behavior of the system under $(+,+)$ boundary conditions. 
For such a system the finite-size scaling theory \cite{BDT2000,Ba83,P90} predicts:
\begin{itemize}
\item For the magnetization (order parameter) profile
\begin{equation}\label{mfss}
m(z,T,h,L)\equiv -\frac{\partial(\beta f)}{\partial h}\simeq a_h L^{-\beta/\nu}X_m
\left(z/L,x_t,x_h\right)
\end{equation}
where
\begin{equation}
x_t=a_t \tau L^{1/\nu}, \qquad x_h=a_h h L^{\Delta/\nu}.
\label{scalingvariables}
\end{equation}

\item For the local (layer) susceptibility profile 
\begin{equation}\label{chifssprofile}
k_B T \, \chi_l(z,T,h,L)\equiv\frac{\partial }{\partial h}m(z,T,h,L)\simeq a_h^2 L^{\gamma/\nu}
X_\chi\left(z/L,x_t,x_h\right)
\end{equation}
with 
$X_\chi\left(z/L,x_\tau ,x_h\right)=\frac{\partial }{\partial x_h}X_m\left(z/L,x_\tau ,x_h\right)$;
\item For the total susceptibility $\chi(T,h,L)\equiv L^{-1}\int_{0}^{L}\chi_l(z,T,h,L)dz$ one has 
\begin{equation}\label{chifss}
k_B T \, \chi(T,h,L)\simeq a_h^2 L^{\gamma/\nu}
X\left(x_t,x_h\right).
\end{equation}
\end{itemize}
In Eqs. (\ref{mfss}) -- (\ref{chifss}), $ \beta $ and $ \gamma $ are the critical exponents for the order parameter and the susceptibility (compressibility), the quantities $a_t$ and $a_h$ are nonuniversal metric factors that can be fixed, for a given system, by taking them to be, e.g., $a_t=1/\left[\xi_0^+\right]^{1/\nu}$, and $a_h=1/\left[\xi_{0,h}\right]^{\Delta/\nu}$. Since the  Ising system with a film geometry $\infty^2 \times L$ possesses a critical point $T_{c,L}$ of its own with coordinates $(x_t^{(c)},x_h^{(c)})$ the scaling functions $X_m,X_\chi$ and $X$ will exhibit singularities near this point. For example 
\begin{equation}
\label{eq:sing_in_X}
X(x_t,x_h^{(c)})\simeq X_{c,t} \;\left( x_t-x_t^{(c)} \right)^{-\gamma_2}, \qquad x_t\to x_t^{(c)},
\end{equation} 
where the subscript in $\gamma_2$ reminds that $\gamma_2$ is the critical exponent of the two-dimensional infinite system that is to be distinguished from the corresponding exponent $\gamma$ for the three dimensional bulk system. 

In the current article we will derive new exact analytical results for the scaling functions $X_m, X_\chi$ and $X$ for the Ginzburg-Landau Ising type mean-field model.   Let us recall that in the mean-field approximation $\beta=\nu=1/2$, $\Delta=3/2$ and $\gamma=\gamma_2=1$. For the version of the model considered here $\xi_0^+=1$ and $\xi_{0,h}=1/\sqrt[3]{3}$ \cite{SHD2003,DSD2007,PHA91}.

We begin our study by presenting a short definition of the model that will help us to introduce some of the notations used further in the article. 

\section{The Ginzburg-Landau mean-field model and basic expressions defining the finite-size quantities of the system}

Let us consider an Ising type critical system in a parallel plate geometry described by the standard $\phi^4$ Ginzburg-Landau Hamiltonian
\begin{equation} \label{FviafIsing}
{\cal F}\left[\phi;\tau,h,L\right] =\int_0^L {\cal L}(\phi,\phi') dz,
\end{equation}
where
\begin{equation}\label{fdefIsing}
{\cal L}\equiv {\cal L}(\phi,\phi')=\frac{1}{2}  {\phi'}^2 +
\frac{1}{2}\tau\phi^2+\frac{1}{4}g\phi^4-h \phi.
\end{equation}
Here: $L$ is the film thickness, ${\phi}$ is the order parameter at the perpendicular position $z$ $(0 < z < L)$, $\tau=(T-T_c)/T_c$ is the bare reduced temperature with $\tau=0$ defining the bulk critical temperature, $h$ is the external ordering field, $g$ is the bare coupling constant and the primes indicate differentiation with respect to the variable $z$. 
Normally, one adds to the right-hand side of Eq. (\ref{FviafIsing}) a surface-type term with parameters used to impose the boundary conditions on the system. We will do that by simply requiring the behavior of the order parameter at the boundaries of the system at $z=0$ and $z=L$ to be of a given prescribed type.

The extrema of the functional ${\cal F}$ are determined by the solutions of the corresponding Euler-Lagrange equation
\begin{equation}\label{Lagrange}
\frac{d}{dz} \frac{\partial {\cal L}}{\partial \phi'}-\frac{\partial {\cal L}}{\partial\phi}=0,
\end{equation}
which, on account of Eq. (\ref{fdefIsing}), reads
\begin{equation}\label{Phieqalone}
{\phi''}-\phi\left[\tau+g\,\phi^2\right]+h=0.
\end{equation}
Multiplying the above equation by $\phi'$ and integrating over $z$ one obtains the following first integral of the system
\begin{equation}\label{IsingStress}
\frac{1}{2}  {\phi'}^2 - 
\frac{1}{2}\tau\phi^2-\frac{1}{4}g\phi^4+h \phi = c,
\end{equation}
where $c$ is the constant of integration.

In the present article we choose the so-called $(+,+)$ boundary conditions:
$\left. \lim \phi\left( z \right) \right| _{z \rightarrow 0}=\left.
\lim \phi \left( z \right) \right| _{z \rightarrow L}=+\infty $. 
Due to the symmetry, under such boundary conditions one shall have that $\phi'(L/2)=0$. It is easy to determine the general behavior of $\phi(z)$ near the boundaries for any fixed finite values of $\tau$ and $h$. Since $\phi(z)\to \infty$ near the boundaries then, say, for the left boundary, Eq. (\ref{IsingStress}) becomes 
\begin{equation}\label{near_boundary}
\phi'\simeq -\sqrt{\frac{g}{2}}\phi^2.
\end{equation}
Solving this equation leads to
\begin{equation}
\label{OP_near_z0}
\phi(z)\simeq\frac{1}{\sqrt{{g/2}} \left|z-z_0\right|},
\end{equation}
where $z_0$ is the position of the boundary (in the case considered $z_0=0$). Note that this leading behavior of the order parameter profile near the boundary does not depend neither on $\tau$, nor on $h$.
From Eq. (\ref{OP_near_z0}) it follows that the integral in Eq.  (\ref{FviafIsing}) diverges. Thus, when a comparison of the values of ${\cal F}$ for different states of the system is needed
either some cut-off of the system near the boundaries is necessary, or 
some special procedure shall be devised. 

From Eq. (\ref{Phieqalone}), using the definition for the local layered susceptibility
\begin{equation}\label{localsusdefmt}
\chi_l(z|\tau,h)\equiv \frac{\partial \phi(z|\tau,h)}{\partial h},
\end{equation}
one obtains that $\chi_l$ fulfills the relation
\begin{equation} \label{eqcontsrchimt}
-\chi_l''+(\tau+ 3g \,\phi^2) \chi_l =1.
\end{equation}
On general grounds, as well as from Eq. (\ref{OP_near_z0}), for a system with $(+,+)$ boundary conditions one immediately obtains  $\chi_l(z\to z_0|\tau,h)=0.$ Combining Eqs. (\ref{OP_near_z0}) and (\ref{eqcontsrchimt}) leads to $\chi_l(z|\tau,h)\to (z-z_0)^2/4$ when $z\to z_0$.

\section{Analytical results for the scaling behavior of the order parameter profiles} 

In terms of the scaling variables
\begin{equation}\label{isvar1}
\zeta=z/L,\qquad x_t=\tau L^{1/\nu}, \qquad
\bar{x}_h=\sqrt{2g} hL^{\Delta/\nu},
\end{equation}
\begin{equation}\label{isvar2}
\phi(z)=\sqrt{\frac{2}{g}}\,L^{-\beta/\nu}X_m(\zeta|x_t,\bar{x}_h),
\end{equation}
with $\beta=\nu=1/2$ and $\Delta=3/2$,
Eq. (\ref{Phieqalone}) for the order parameter profile takes the form
\begin{equation}
\label{msaclingEqA}
X_m''(\zeta)=X_m(\zeta)\left[ x_t+2X_m^2(\zeta)\right]-\frac{\bar{x}_h}{2}.
\end{equation}
The solutions of \eq{msaclingEqA} determine the extrema of the energy functional
\begin{equation} \label{afunctional}
{\cal E} =\int_0^1 f(X_m(\zeta),X_m'(\zeta)) d\zeta, 
\end{equation}
where
 \begin{equation}\label{fdefIsingmScalingA}
f(X_m(\zeta),X_m'(\zeta))=\left[{X_m'(\zeta)}\right]^2 + X_m^4(\zeta)+
x_t X_m^2(\zeta)-\bar{x}_h X_m(\zeta)
\end{equation}
is the energy density. Hereafter, the primes indicate differentiation with respect to the variable $\zeta$ which, as follows from Eq. (\ref{isvar1}), varies in the closed interval $[0,1]$.
According to Eq. (\ref{IsingStress}) the first integral of Eq. (\ref{msaclingEqA})  reads
 \begin{equation}\label{FirstIntA}
\left[{X_m'(\zeta)}\right]^2 =P[X_m], \qquad
P[X_m]=X_m^4(\zeta)+x_t X_m^2(\zeta) -\bar{x}_h X_m(\zeta)+\varepsilon,
\end{equation}
where $\varepsilon$ denotes the respective constant of integration.

\subsection{Analytical representation of the order parameter profiles in the case of zero field} 

When $h=0$ the magnetization profile  is known exactly
\cite{K97} in terms of two mutually related implicit equations:
\begin{enumerate}

\item[a)] when $x_t\equiv \tau L^2\ge -\pi^2$
\begin{eqnarray}\label{mz}
X_m(\zeta|x_t,0) &=&2 K(k)\frac{{\rm dn}[2 K(k)\zeta;k]}{{\rm sn}[2 K(k)\zeta;k]},
\end{eqnarray}
where $k^2 \ge 0$ is to be determined from
\begin{equation}\label{tk}
    x_t=[2K(k)]^2(2k^2-1);
\end{equation}

\item[b)] when $x_t \le -\pi^2$
\begin{eqnarray}\label{mzd}
X_m(\zeta|x_t,0) &=& \frac{2 K(\bar{k})}{{\rm sn}[2 K(\bar{k})\zeta;\bar{k}]},
\end{eqnarray}
where $\bar{k}^2\ge 0$ is to be determined from
\begin{equation}\label{tkb}
    x_t=-[2K(\bar{k})]^2(\bar{k}^2+1).
\end{equation}
\end{enumerate}
Here $K(k)$ is the complete elliptic integral of the first kind,
${\rm dn}(\zeta;k)$ and ${\rm sn}(\zeta;k)$ are the Jacobian
delta amplitude and the sine amplitude functions, respectively.
The bulk critical point $T=T_c$ corresponds to $k^2=1/2$.
%The above expressions are consistent with the following scaling form for the order parameter:
%\begin{equation}\label{sf}
%\phi(z)=L^{-\beta/\nu}X_\phi\left(z/L,tL^{1/\nu}\right),
%\end{equation}
%with $\beta=\nu=1/2$.
Note, however, that within the mean-field theory the magnitude of the variable $\phi$ is not universal, in that it is multiplied by the nonuniversal factor $\sqrt{2/g}$ -- see Eq. (\ref{isvar2}). Finally, we stress that the choice of two parameterizations (see Eqs. (\ref{tk}) and (\ref{tkb})) of the scaling functions in Eqs. (\ref{mz}) and (\ref{mzd}), is just for convenience; it allows one to avoid using imaginary values of $k$ and $\bar{k}$. Indeed, one can transfer any of the set of equations into the other one. For example,  defining $\bar{k}$ as $\bar{k}=i {k}/ {k'}$
%\begin{equation}\label{krel}
%\bar{k}=i \frac{k}{k'}, \qquad 
%\end{equation}
where $k'^{\,2}=1-k^2 $,
and taking into account the following properties of the elliptic functions \cite{GR,AS} $K(\bar{k})=k'\; K(k)$
%\begin{equation}\label{kKrel}
%K(\bar{k})=k'\; K(k),
%\end{equation}
and
\begin{equation}\label{dnsnrel}
\frac{{\rm dn}(u;ik)}{{\rm sn}(u;ik}=\frac{\sqrt{1+k^2}}{{\rm sn}(u \sqrt{1+k^2};k/\sqrt{1+k^2})},
\end{equation}
one can easily check that the pair of equations (\ref{mz}), (\ref{tk}) is equivalent to the pair of equations (\ref{mzd}), (\ref{tkb}).

In order to utilize the symmetry of the problem it is helpful to move the coordinate frame so that the origin of the $\zeta$ axis is at the midpoint of the film \cite{DRB2009}. Taking into account that \cite{GR}
\begin{equation}\label{elfunperprop}
\frac{{\rm dn}[u+K(k);k]}{{\rm sn}[u+K(k);k]}=\frac{k'}{{\rm cn}(u;k)}, \qquad \mbox{and} \qquad {\rm cn}(iu;k')=1/{\rm cn}(u;k) 
\end{equation}
%and that
%\begin{equation}\label{elfuncnprop}
%{\rm cn}(iu;k')=1/{\rm cn}(u;k),
%\end{equation}
from Eq. (\ref{mz}) one  obtains \cite{DRB2009}
\begin{equation}\label{Xmmiddle}
X_m(\zeta|x_t)=X_{m0} \;{\rm cn}[i 2K(k) \zeta;k']
\end{equation}
where $\zeta\in [-1/2,1/2]$ and
\begin{equation}\label{Xm0def}
X_{m0} \equiv 2 k' K(k)=2 K(\bar{k}).
\end{equation}
Since ${\rm cn}(0;k)=1$, one has $X_{m0}=X_m(0|x_t)$.
%\begin{equation}\label{Xm0}
%X_{m0}=X_m(0|x_t).
%\end{equation}
Eq. (\ref{Xmmiddle}) is the simplest representation of the order parameter profile in a system with strongly adsorbing boundaries we are aware of. 
\begin{figure}[htbp]
\centering
\includegraphics[width=3in]{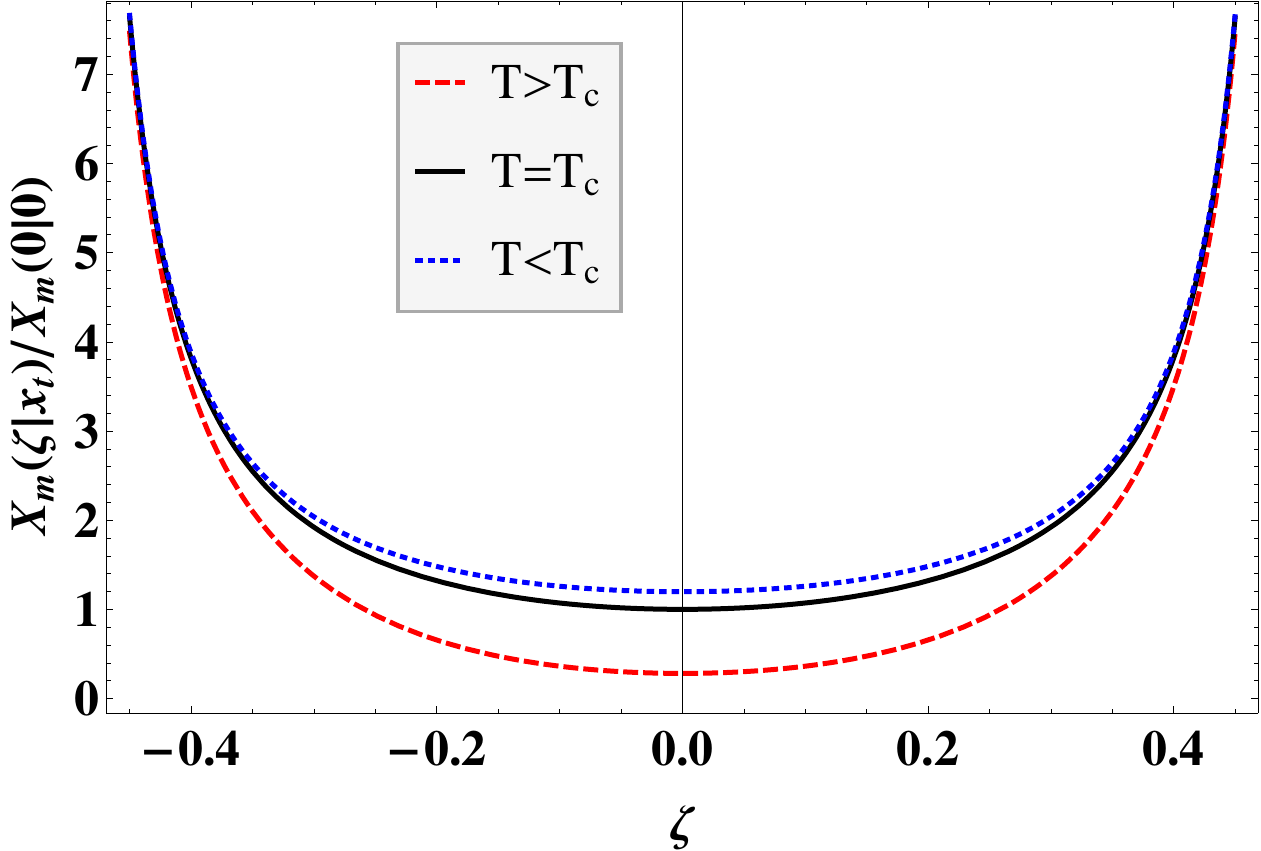}
\caption{Color online. Plot of the scaling  function of the order parameter profile $X_m(\zeta|x_t)$ -- see Eq. (\ref{Xmmiddle}), normalized per its value in the middle of the system $X_m(\zeta=0|x_t=0)$ for $T=T_c$.}
\label{fig:Xm}
\end{figure}

The typical behavior of $X_m(\zeta|x_t)$ is shown on figure \ref{fig:Xm} for three different temperatures: at, well below, and well above the bulk critical temperature $T_c$.

\subsection{Analytical results for the order parameter profiles in the case of nonzero field} 

When $h \neq 0$ the study of the order parameter (magnetization) profiles was carried out numerically \cite{SHD2003}. Below, however, we shall give an analytical representation of the foregoing profiles by means of Weierstrass elliptic functions, which is quite similar to that presented above for $h=0$ up to the lack of the
parametrization through the elliptic modulus achieved in that case,
see Eqs. (\ref{mz}) -- (\ref{tkb}).

First, let us recall that
%in the present work
we are interested in real-valued solutions $X_m(\zeta)$ of Eq. (\ref{msaclingEqA}), corresponding to given values of the parameters $x_t$ and  $\bar{x}_h$, which are smooth in the open interval $(0,1)$ and satisfy the $(+,+)$ boundary conditions. 
Evidently, for each such solution the polynomial $P[X_m]$ should have at least one real root $X_{m0}$. Otherwise the considered solution would be either strictly increasing or strictly decreasing, contrary to the required boundary conditions, since its derivative $X_m'(\zeta)$ would be either strictly positive or strictly negative as implied by the particular form of Eq.~(\ref{FirstIntA}).
Consequently, the constant of integration $ \varepsilon $ corresponding to such a solution can be cast in the form
\begin{equation}\label{IntConst}
\varepsilon=-X_{m0}\left(X_{m0}^{3}+x_{t}X_{m0}-\bar{x}_{h}\right).
\end{equation}
Now, given a triple of values of the parameters $x_t$, $\bar{x}_h$ and $X_{m0}$, each real-valued solution $X_m(\zeta)$ of Eq. (\ref{FirstIntA}) can be expressed, following \cite[\S20.22, \S21.73]{WW63}, 
%, up to an arbitrary translation of the independent variable $ \zeta $
in the form
\begin{equation}  \label{WSA}
X_m \left( \zeta| x_t,\bar{x}_{h},X_{m0}\right) =X_{m0}+\frac{6 X_{m0} \left(x_t+2 X_{m0}^2\right)- 3 \bar{x}_{h}}{12 \wp \left( \zeta - \frac{1}{2};g_{2},g_{3}\right) - \left(x_t+6 X_{m0}^2\right)} \cdot
\end{equation}
Here, $\wp \left(\xi;g_{2},g_{3}\right)$ is the Weierstrass elliptic function, $g_{2}$ and $g_{3}$ are the invariants of the polynomial $P[X_m]$, which according to \cite[\S20.22, \S21.73]{WW63} and Eq. (\ref{IntConst}) read
\begin{equation} \label{g1} 
g_2=\frac{1}{12} x_t^2-X_{m0}\left(X_{m0}^{3} +x_{t}X_{m0}-\bar{x}_{h}\right),
\end{equation}
\begin{equation} \label{g2} 
g_3=-\frac{1}{432} \left[27 \bar{x}_{h}^2+2 x_t^3+72 x_t X_{m0}\left( X_{m0}^{3}+x_{t}X_{m0}-\bar{x}_{h}\right)\right].
\end{equation}

Note that each function of form (\ref{WSA}) has the following important properties.
First, its graph in the $ (\zeta, X_{m}) $ plane is symmetric with respect to the line  parallel to $X_{m}$ axis and passing through the point $ (1/2, 0) $. This is because the Weierstrass elliptic functions have the property $ \wp \left(-\xi;g_{2},g_{3}\right)=\wp \left(\xi;g_{2},g_{3}\right) $.
Next,
\begin{equation} \label{Pr1} 
X_m \left( \frac{1}{2}| x_t,\bar{x}_{h},X_{m0}\right) =X_{m0}, \qquad
X_m' \left( \frac{1}{2}| x_t,\bar{x}_{h},X_{m0}\right) =0,
\end{equation}
since the function $  \wp \left( \zeta - 1/2;g_{2},g_{3}\right) $ has a second-order pole at $ \zeta = 1/2$, i.e., $\left. \lim  \wp \left( \zeta - 1/2;g_{2},g_{3}\right) \right| _{\zeta \rightarrow 1/2}=+\infty $.
Finally, the considered function has a local minimum at $ \zeta = 1/2$ if and only if 
\begin{equation}
\label{eq:condition}
2 X_{m0} \left(x_t+2 X_{m0}^2\right)- \bar{x}_{h}>0
\end{equation}
since Eqs. (\ref{msaclingEqA}) and (\ref{Pr1}) imply 
\begin{equation}
X_m''\left( \frac{1}{2}| x_t,\bar{x}_{h},X_{m0}\right)=
\frac{1}{2}\left[ 2 X_{m0} \left(x_t+2 X_{m0}^2\right)- \bar{x}_{h}\right].
\end{equation}

Let us stress that not any function of the form (\ref{WSA}) satisfies the required $ (+,+) $ boundary conditions.
Actually, a function of form (\ref{WSA}) corresponding to a given couple of values of the parameters $x_t$ and  $\bar{x}_h$ meets these conditions if and only if $ X_{m0}$ is such that:

\noindent {\it (a)} the denominator of the second term in the right hand side of expression (\ref{WSA}) attains zero at $\zeta=0 $ and $\zeta=1$, which in view of the relation
$\wp \left(-\xi;g_{2},g_{3}\right)=\wp \left(\xi;g_{2},g_{3}\right)$,
means 
\begin{equation}
\label{TrEq_text}
\varphi(x_t, \bar{x}_h, X_{m0})=0,
\end{equation}
where
\begin{equation}
\label{TrEq_text_1}
\varphi(x_t, \bar{x}_h, X_{m0}):=12 \wp \left( \frac{1}{2};g_{2},g_{3}\right) - \left(x_t+6 X_{m0}^2\right);
\end{equation}
{\it (b)} $X_{m0}$ satisfies constraint (\ref{eq:condition}), meaning that the function $ X_m \left( \zeta| x_t,\bar{x}_{h},X_{m0}\right) $ attains its minimal value at $ \zeta = 1/2$, i.e., at the center of the system. 

Thus, in order to determine the smooth for each $\zeta \in (0,1)$  functions $ X_m \left( \zeta| x_t,\bar{x}_{h},X_{m0}\right) $ of form (\ref{WSA}) satisfying the considered  $(+,+)$ boundary conditions for given values of the parameters $x_t$ and $\bar{x}_{h}$ one should find all the solutions $X_{m0}$ of the respective transcendental equation (\ref{TrEq_text}), which are such that constraint (\ref{eq:condition}) is fulfilled. Any such function will represent an order parameter profile satisfying the $(+,+)$ boundary conditions. In the cases in which the parameters $x_t$ and $\bar{x}_{h}$ are such that there is more than one value of the parameter $ X_{m0}$ satisfying the above requirements, i.e., there is more than one order parameter profile satisfying the $(+,+)$  boundary conditions, on physical grounds we chose the one that minimizes the  truncated energy
\begin{equation} \label{tfunctional}
{\cal E}_{tr} (x_t, \bar{x}_{h}, X_{m0})=2 \int_\delta^{\frac{1}{2}} f(X_m(\zeta),X_m'(\zeta)) d\zeta,
\end{equation}
were $ \delta $ is a small positive number, i.e., $ \delta \ll 1$. We work with the truncated, instead with the full energy, since the integral (\ref{afunctional}) determining the energy $ \cal E $ of the states of the system is divergent. Below we justify this procedure in a mathematically rigorous way. 
Before proceeding to that, let us introduce an approximation of the considered order parameter profiles near the singular point $\zeta=0$.

Given $x_t$, $\bar{x}_h$ and $\varepsilon=-X_{m0}\left(X_{m0}^{3}+x_{t}X_{m0}-\bar{x}_{h}\right)$, the solution of Eq. (\ref{FirstIntA}), which is unique, can be approximated near the singular point $\zeta=0$ by the function
\begin{equation}
\label{FIsmall_z_behavior_of_Xm}
\tilde{X}_m(\zeta)= \frac{1}{\zeta}-\frac{x_t}{6} \zeta+\frac{\bar{x}_h }{8}\zeta^2
+\frac{7 x_t ^2-36 \varepsilon}{360 } \zeta^3
-\frac{\bar{x}_h x_t }{48} \zeta^4.
\end{equation}
Indeed, the substitution $X_m=\tilde{X}_m$ in Eq. (\ref{FirstIntA}) gives
\begin{equation}\label{FirstIntXapprox}
\fl \left[\tilde{X}_m'(\zeta)\right]^2 -
\tilde{X}_m^4(\zeta)-x_t \tilde{X}_m^2(\zeta) +\bar{x}_h \tilde{X}_m(\zeta)-\varepsilon=
\left(\frac{\varepsilon x_t }{10}+\frac{3\bar{x}_h^2}{32}-\frac{31 x_t ^3}{1080}\right) \zeta^2+{\cal O}(\zeta^ {3}).
\end{equation}
Consequently, using the approximate solution $\tilde{X}_m$ of Eq. (\ref{FirstIntA}), given by Eq. (\ref{FIsmall_z_behavior_of_Xm}), one can approximate the energy density (\ref{fdefIsingmScalingA}) near the singular point $\zeta=0$ as 
\begin{equation} \label{ApproxED}
f(\tilde{X}_m (\zeta),\tilde{X}_m'(\zeta))=\frac{2}{\zeta^4}+\frac{2 x_t }{3 \zeta^2}-\frac{\bar{x}_h}{\zeta}-\frac{8 x_t ^2}{45}+\frac{\varepsilon}{5}+
{\cal O}(\zeta).
\end{equation}
Next, using the symmetry of the energy density with respect to the point $ \zeta=1/2 $, implied by the symmetry of the general solution (\ref{WSA}) with respect to this point and Eq. (\ref{fdefIsingmScalingA}), 
we rewrite the energy (\ref{afunctional}) in the form
\begin{equation} \label{TrunkatedEnergy}
{\cal E} (x_t, \bar{x}_{h}, X_{m0})=
2 \int_0^{\delta} f(X_m(\zeta),X_m'(\zeta)) d\zeta+
{\cal E}_{tr} (x_t, \bar{x}_{h}, X_{m0};\delta). 
\end{equation}
Now, let $ X_{m01}$ and $ X_{m02}$ be two different values of the parameter $X_{m0}$ determining two different states of the system for same values of the parameters $x_t$ and  $\bar{x}_{h}$. Then, using near the singular point $\zeta=0$ the approximate energy density (\ref{ApproxED}) corresponding to the approximate solution  (\ref{FIsmall_z_behavior_of_Xm}), we observe that
\begin{eqnarray}\label{eq:energy_fiff}
 \fl \mathcal{E}(x_{t},\bar{x}_{h},X_{m02})- \mathcal{E}(x_{t},\bar{x}_{h},X_{m01})&&= \mathcal{E}_{tr}(x_{t},\bar{x}_{h},X_{m02};\delta)-\,\mathcal{E}_{tr}(x_{t},\bar{x}_{h},X_{m01};\delta) \\
&&+\frac{2}{5}\left[ \varepsilon (x_{t},\bar{x}_{h},X_{m02})-\varepsilon
(x_{t},\bar{x}_{h},X_{m01})\right] \delta 
+\mathcal{O}(\delta ^{2}) \nonumber
\end{eqnarray}
and, hence, the difference between the energies of the two regarded states is well defined and determined by the difference between the two respective truncated energies up to terms of order ${O}(\delta) $.
As it is clear from \eq{eq:energy_fiff}, $\delta$ determines the precision with which  we determine the energy differences between any two solutions of the oder parameter problem.  Since the value of $\delta$ is on our disposal, we can, at least in principle, determine these energy differences to any prescribed precision. A numerical verification of the above relation is presented in the Appendix. 

In the remainder, using the derived exact analytical expressions described above in this Subsection, we study the behavior of the order parameter profiles in the critical  and in the capillary condensation regimes. It should be stressed, however, that the solutions $X_{m0}$ of the transcendental equation (\ref{TrEq_text}) corresponding to given values of the parameters $x_t$ and $\bar{x}_{h}$ are determined numerically and those of them that meet the other necessary conditions are identified by inspection.

\begin{figure}[htbp]
\centering
\includegraphics[width=2.6in]{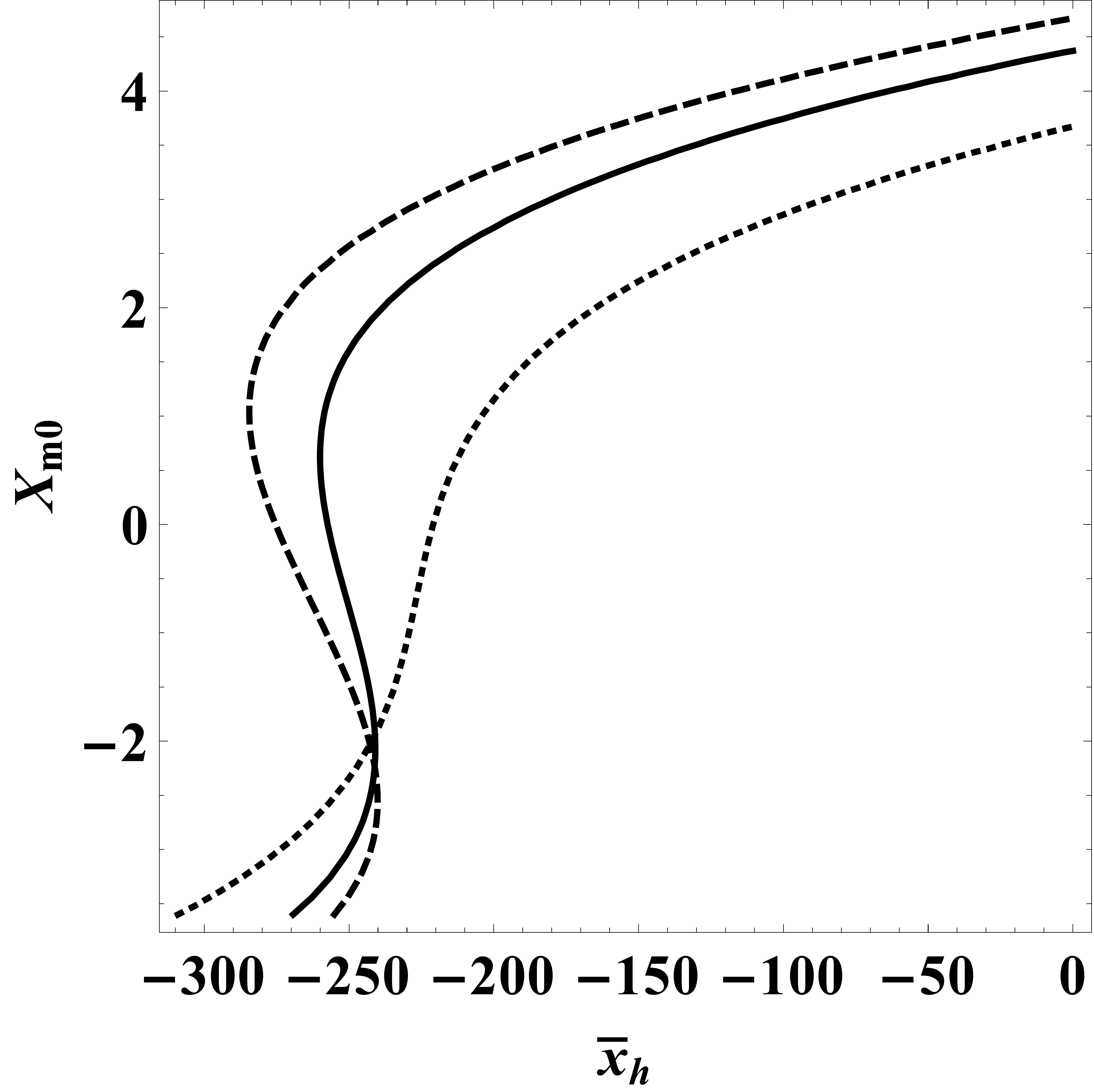} \qquad
\includegraphics[width=2.6in]{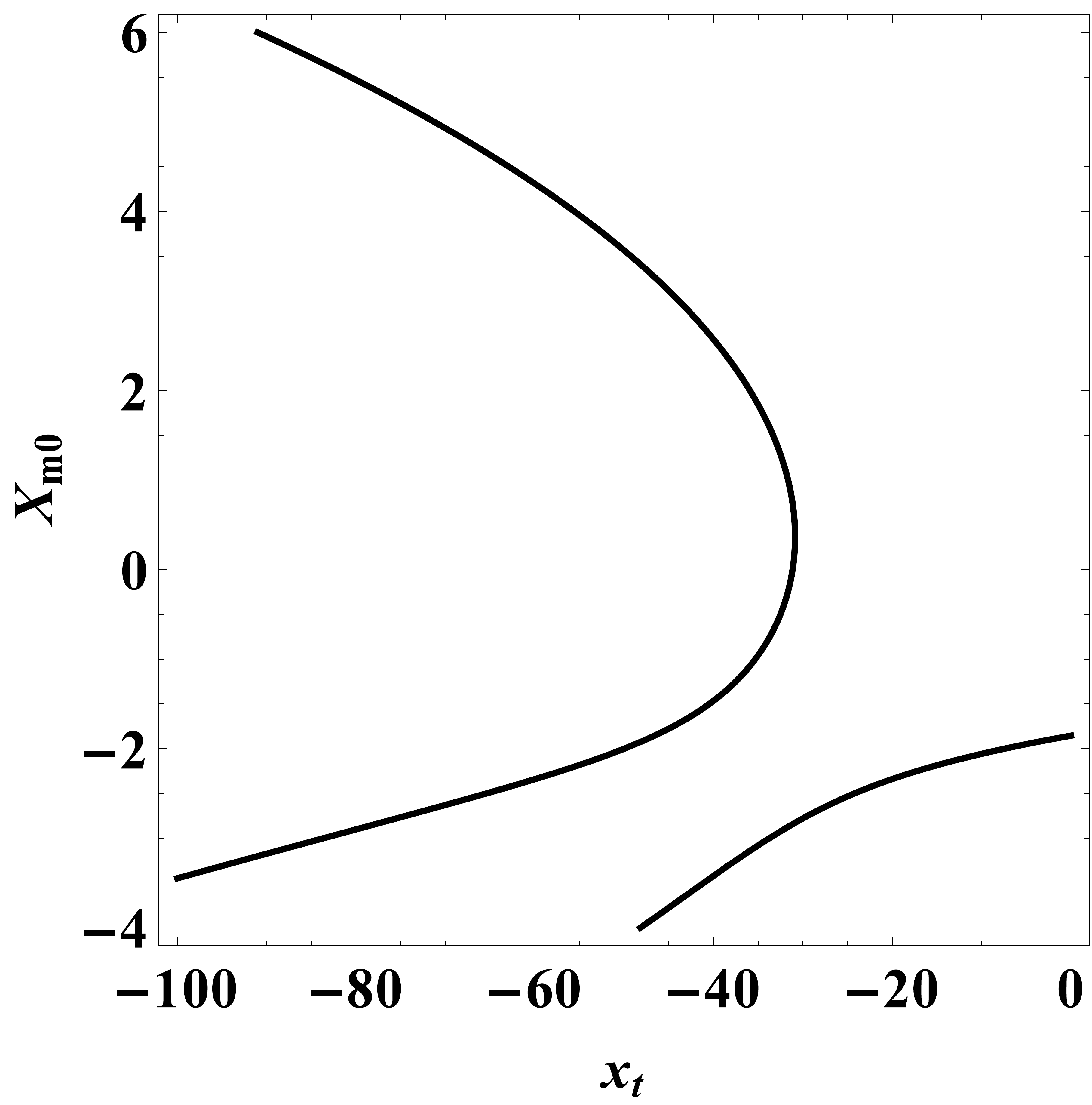}
\caption{The dependence of the value $X_{m0}$ of the order parameter in the middle of the film on $x_t$ and $\bar{x}_{h}$. Left: $x_{t}=-20$ (dotted), $x_{t}=-33.8105$ (thick) and $x_{t}=-40$ (dashed); right: $\bar{x}_{h}=-250$.}
\label{fig:PhD}
\end{figure}

\begin{figure}[htbp]
\centering
\includegraphics[width=2.6in]{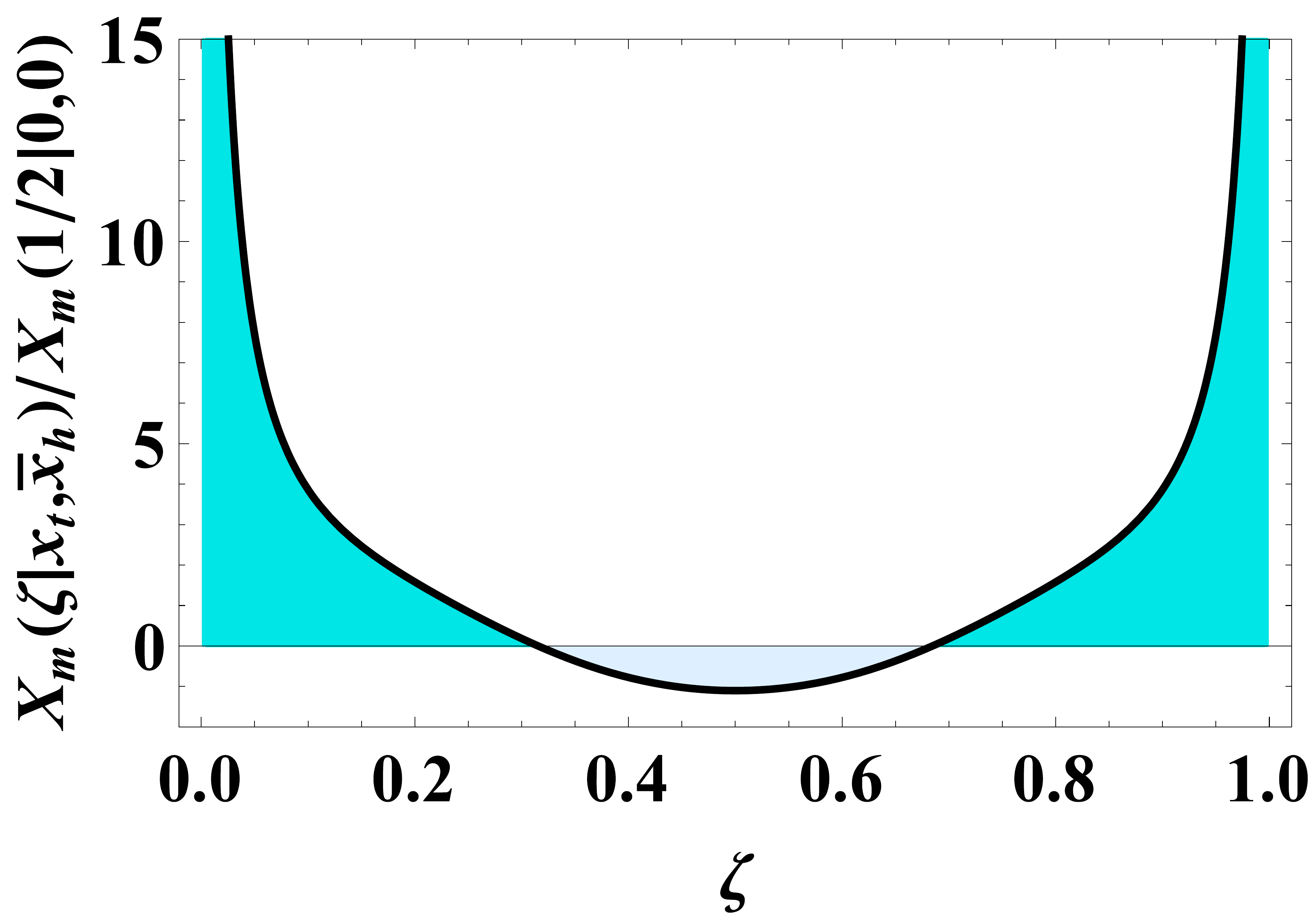} \qquad
\includegraphics[width=2.6in]{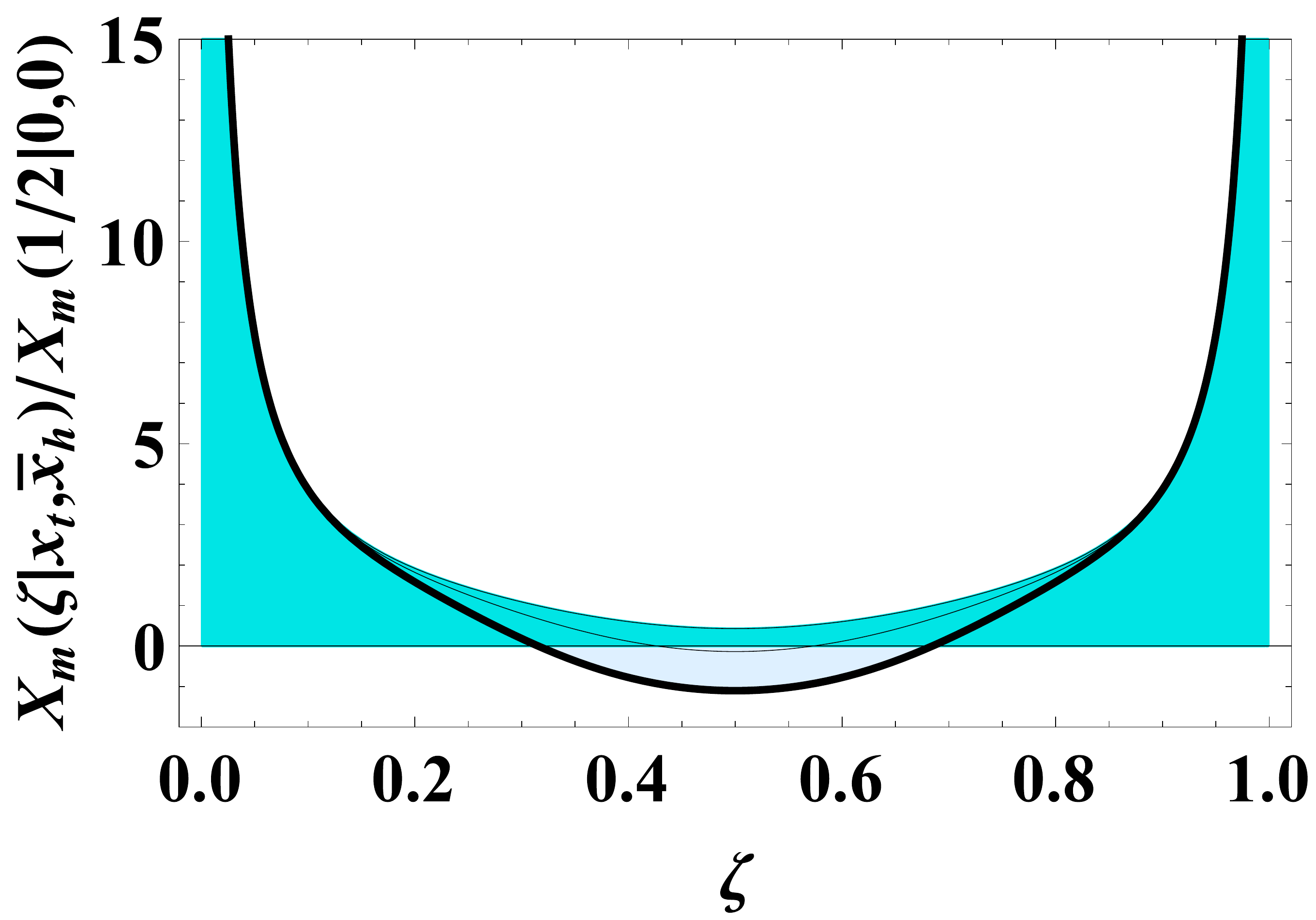}
\caption{Order parameter profiles at $\bar{x}_{h}=-250$ corresponding to $x_{t}=-30$ (left) and $x_{t}=-32$ (right), where the stable profile is represented by the thick curve.}
\label{fig:Ex1}
\end{figure}

Our first observation concerns the number of the solutions of the considered $ (+,+) $ boundary value problem. There are values of the  parameters $x_t$ and $\bar{x}_{h}$ for which we find only one solution of the problem, but there are also regions in the temperature-field plane where there exist three solutions satisfying the $(+,+)$ boundary conditions, which reduce to two in certain limiting cases. These alternatives are illustrated in figures \ref{fig:PhD} and \ref{fig:Ex1}.
In figure \ref{fig:PhD} we show the evolution with $ x_t $ (left) and $\bar{x}_{h}$ (right) of the value $X_{m0}$ of the order parameter in the middle of the film for three different values of the temperature $ x_t $ and for one value of the field $\bar{x}_{h}$, respectively.
The order parameter profiles corresponding to $\bar{x}_{h}=-250$,  $x_{t}=-30$ and $\bar{x}_{h}=-250$, $x_{t}=-32$ are depicted in figure \ref{fig:Ex1}, where the stable profiles are represented by thick curves.


The existence of more than one order parameter profiles for one and the same temperature-field combination is a necessary, but not sufficient, condition for the occurrence of capillary condensation transition. The latter takes place when at least two of the observed order parameter profiles have the same energy meaning that they coexist. 

\begin{figure}[htbp]
\centering
\includegraphics[width=2.9in]{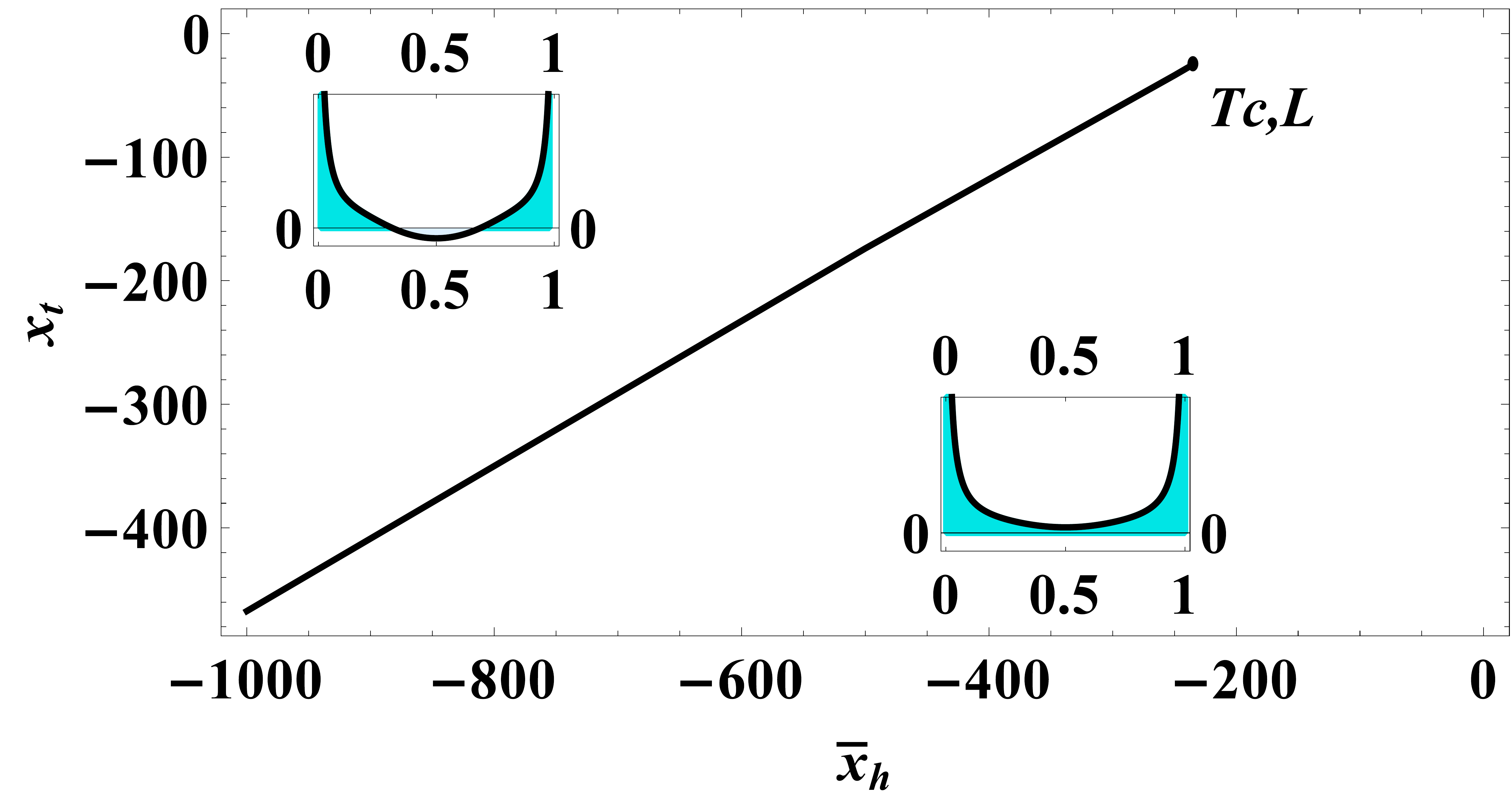} \quad
\includegraphics[width=2.9in]{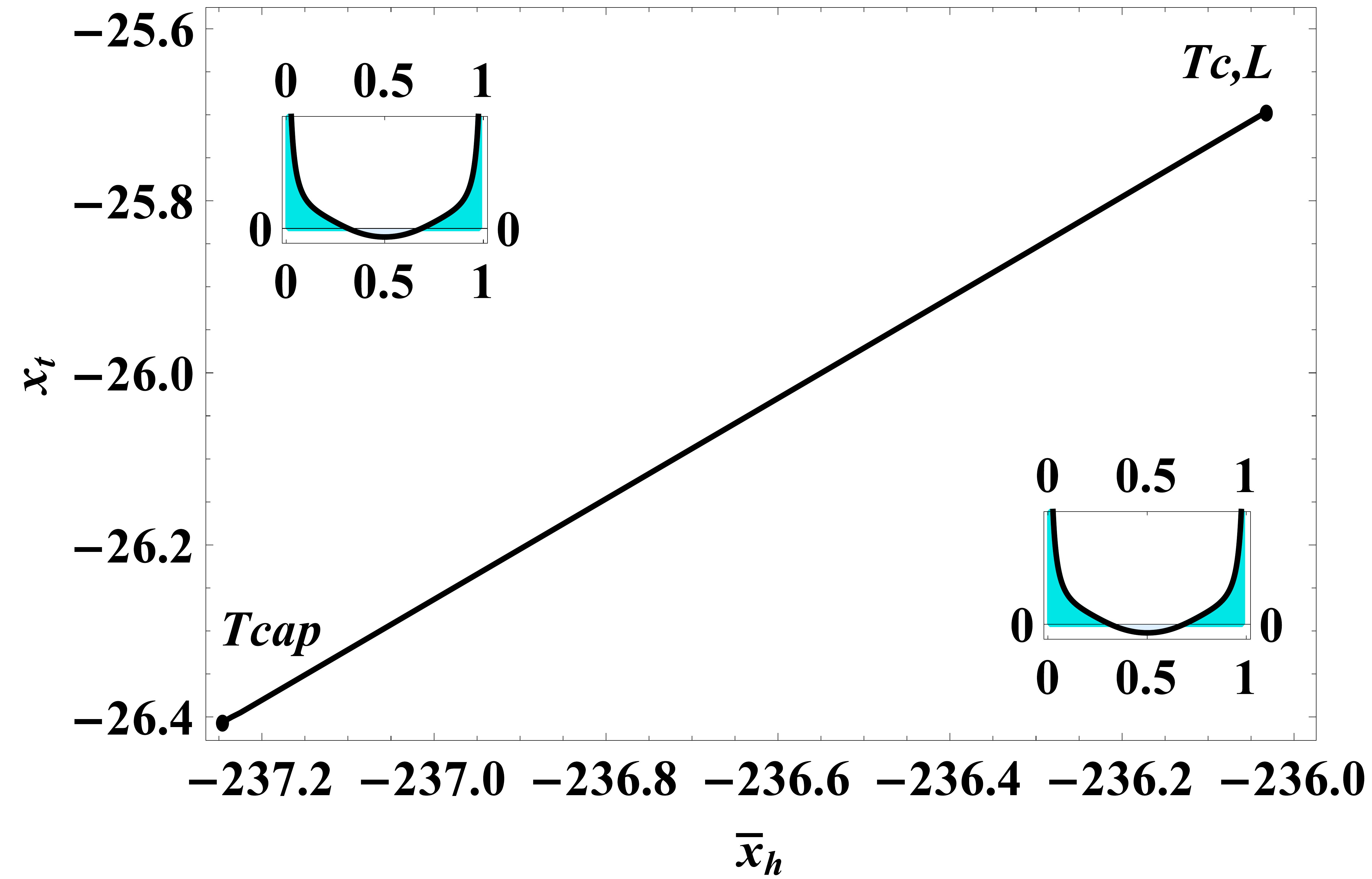}
\caption{Phase diagram (left) showing the border line crossing which the system jumps from a gas to either a denser gas or liquid state in the middle of the system. This curve ends at the critical point of the finite system $T_{c,L}$, $(x_t^{(c)},x_h^{(c)})=(-25.6983391, -236.0350005)$. The figure on the right shows the pre-capillary-condensation curve where above $ T_{\rm cap}$ and below $ T_{c,L}$ the system jumps from a less dense gas to a more dense one. For $T\le T_{\rm cap}$ the order parameter in the middle of the system jumps from a gas to a liquid state.}
\label{fig:DP}
\end{figure}

\begin{figure}[htbp]
\centering
\includegraphics[width=2.6in]{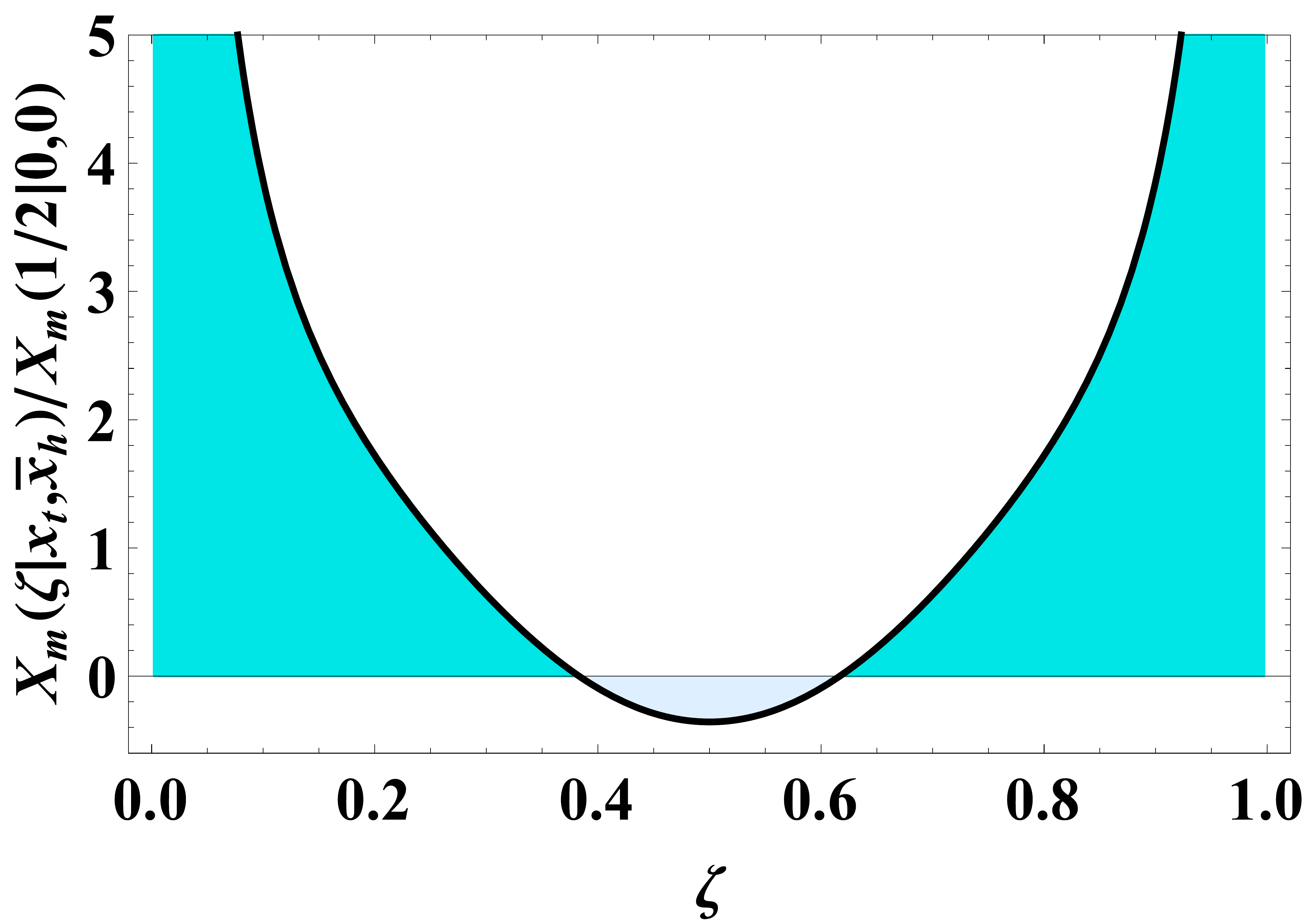} \qquad
\includegraphics[width=2.6in]{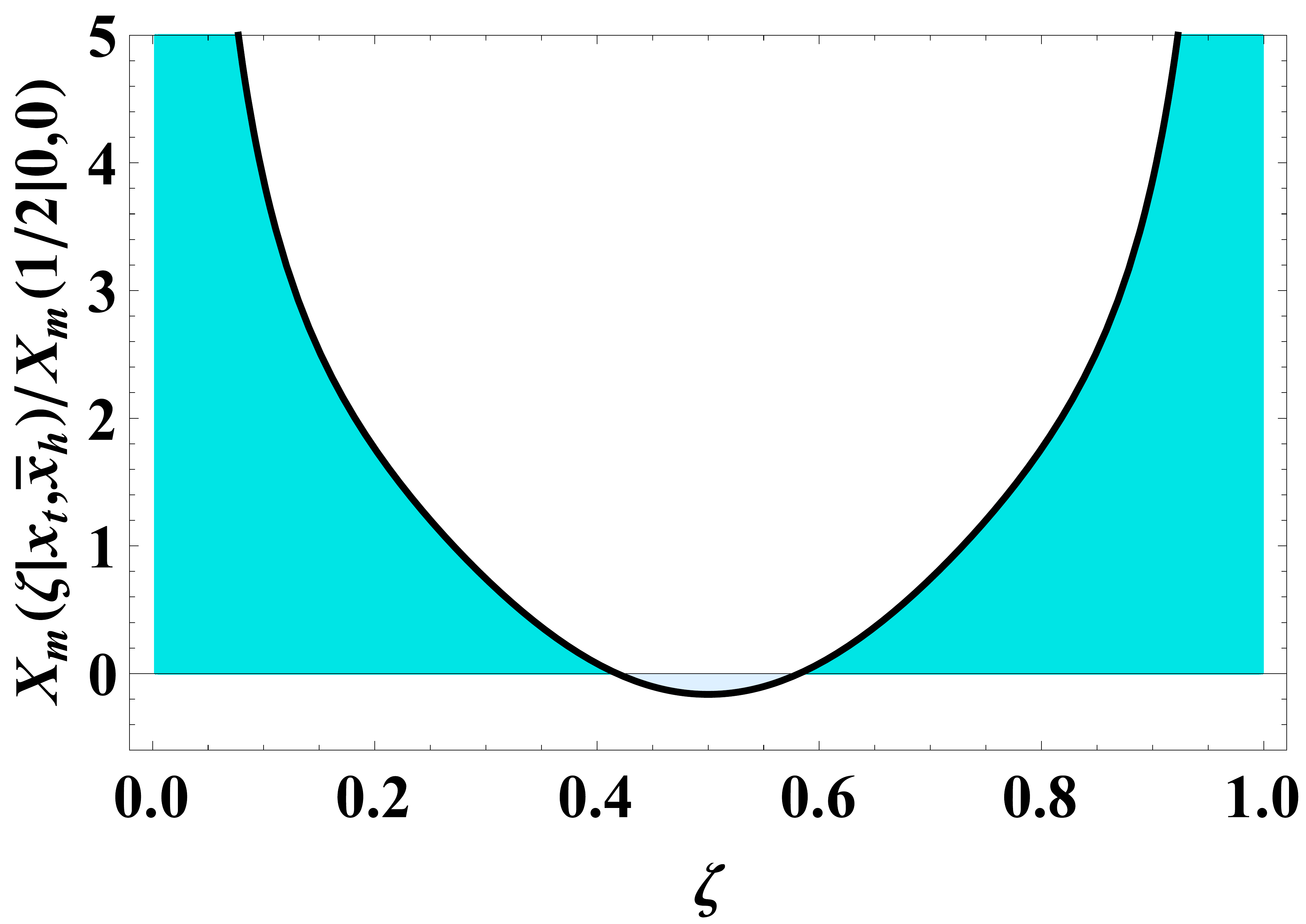}
\caption{The pair of order parameter profiles coexisting at pre-capillary-condensation for $\bar{x}_{h}=-236.2$ and $x_{t}=-25.795$.}
\label{fig:GGOP}
\end{figure}

\begin{figure}[htbp]
\centering
\includegraphics[width=2.7in]{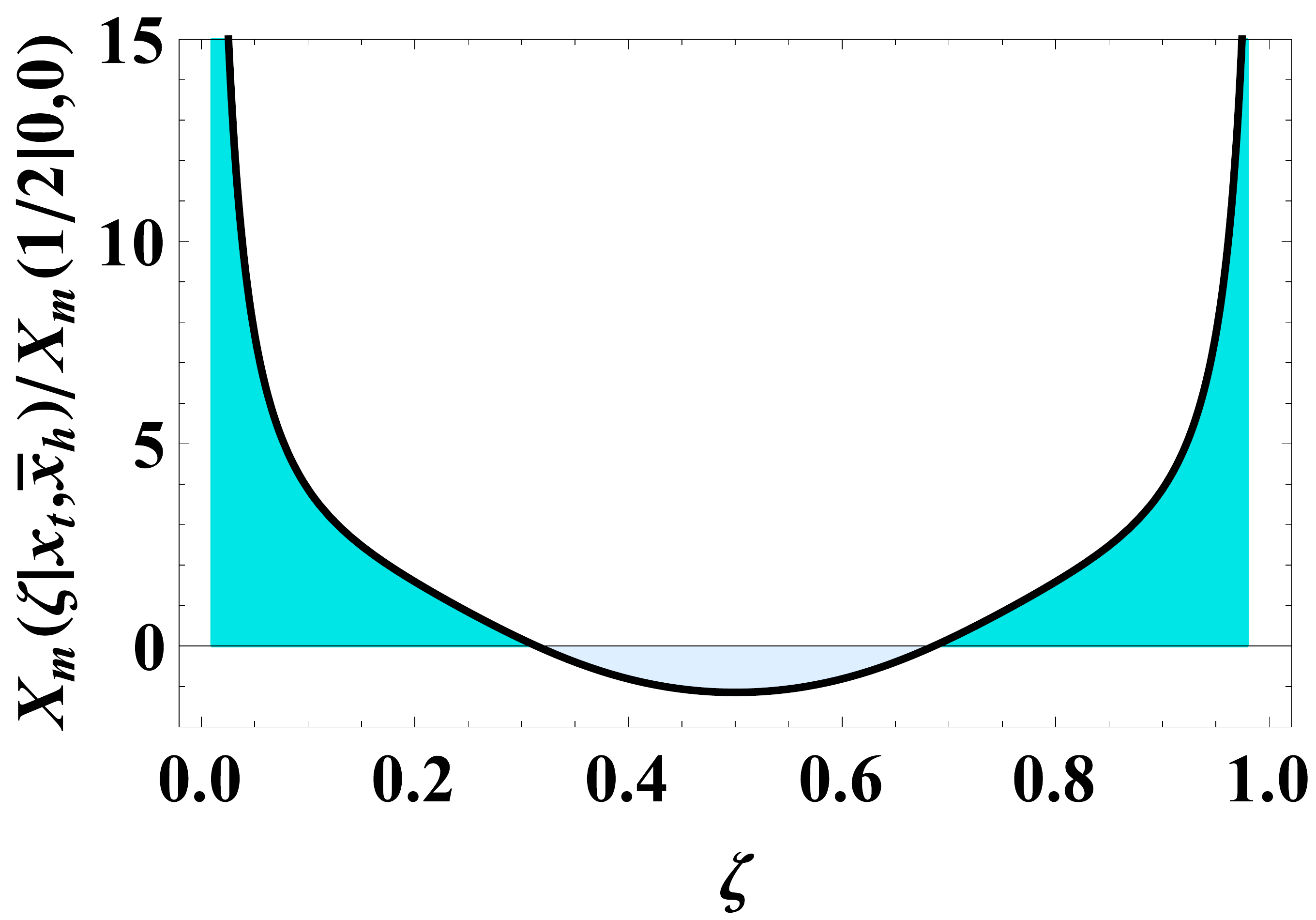}\qquad
\includegraphics[width=2.7in]{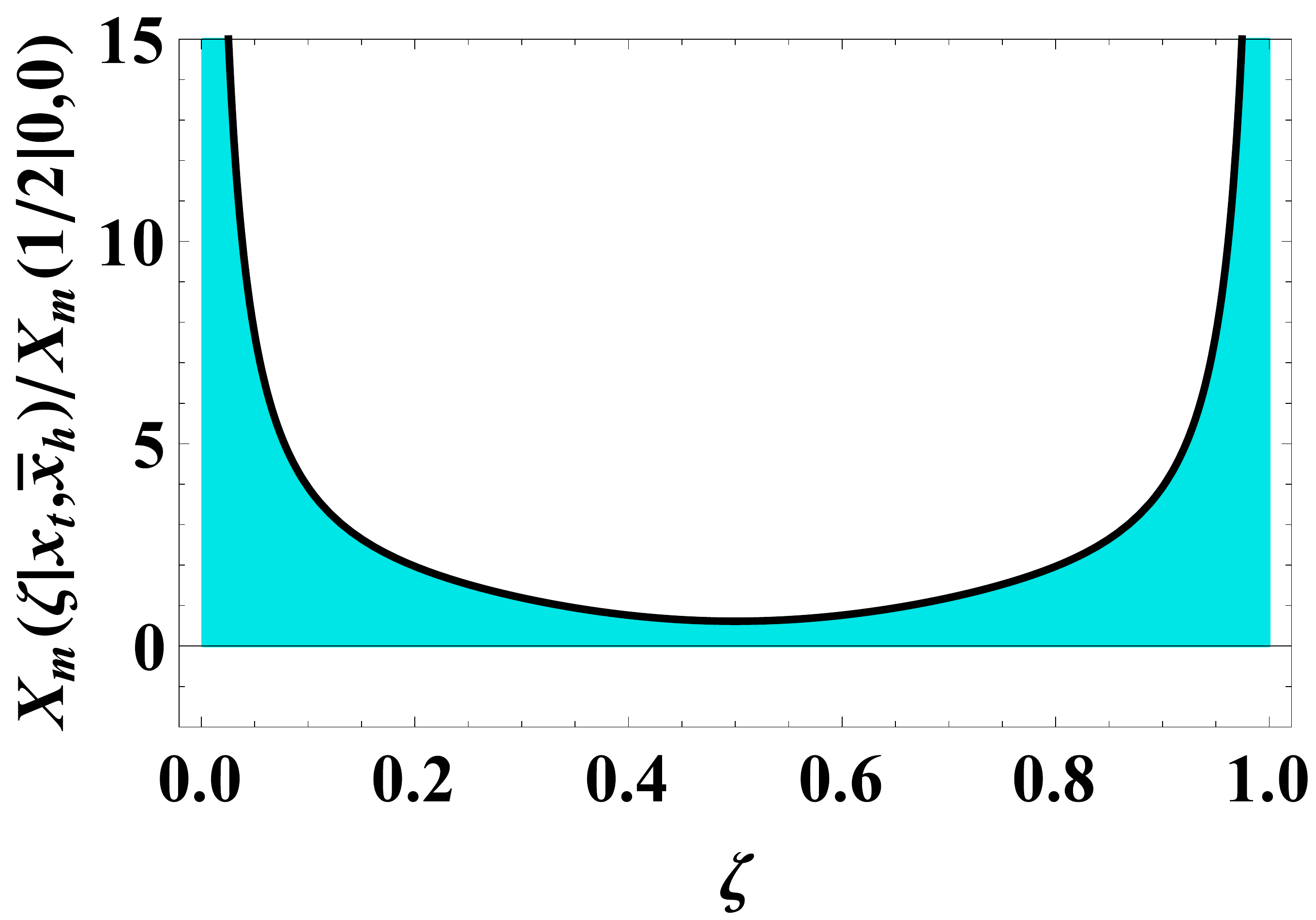}
\caption{The pair of order parameter profiles coexisting at capillary condensation for $\bar{x}_{h}=-250$ and $x_{t}=-33.8105$.}
\label{fig:RE}
\end{figure}

Based on the derived exact analytical expressions we obtain the phase diagram (see figure \ref{fig:DP}, left) and an unexpected result of the existence of a curve in the temperature-field plane (see figure \ref{fig:DP}, right) where the system jumps below its bulk critical temperature from a less dense gas to a more dense gas (see figure \ref{fig:GGOP}) before switching on continuously into the usual jump from gas to liquid state in the middle of the system in the capillary condensation regime  (see figure \ref{fig:RE}). Some technical details related to the determination of the data presented in these plots are given in the Appendix.

We stress here that we term a given state of the system "gas", or "gas-like", if $X_m(1/2|x_t,\bar{x}_h,X_{m0})<0$ and "liquid", or "liquid-like", when $X_m(1/2|x_t,\bar{x}_h,X_{m0})>0$. We remind that under the $(+,+)$ boundary conditions studied in the current article one always has $X_m(\zeta|,x_t,\bar{x}_h,X_{m0})>0$ for $\zeta$ close enough to $0$ or $1$, i.e., one always observe a "liquid-like" state near the boundaries of the system. When one lowers the temperature the following is happening. Above $T_{c,L}$ one has a single order parameter profile that satisfies the $(+,+)$ boundary conditions. Near the phase transitions line  there are already three such profiles two of which provide at a point belonging to the phase line and characterized with given $x_t$ and $x_h$  the minimum of the energy of the system, i.e., they describe the phase coexistence between the gas and the liquid phases. When crossing this line the "liquid-like" order parameter profile changes abruptly with the liquid phase intruding deeper into the capillary. It turns out, however, that in a given temperature range the two liquid branches  stemming from the two surfaces of the capillary do not meet in the middle, but a "gas-lke" gap still exists there for temperatures $T_{\rm cap}<T<T_{c,L}$. We call the phase coexistence line for this special case pre-capillary-condensation curve. Upon further reduction of the temperature or increase of the magnitude of the negative external field the density of the fluid in this gap continuously  increases reaching its liquid value. Thus, if one defines the capillary condensation temperature $T_{\rm cap}$ as the highest one at which the entire capillary fills with liquid one will obtain $T_{\rm cap}$ that differs from $T_{c,L}$ on a scale determined by $L^{-1/\nu}$. Of course, this is a result that follows within the model considered and will be desirable to check if it is a specific feature of the model or if it can be experimentally verified. 

%\begin{figure}[htbp]
%\centering
%\includegraphics[width=2.6in]{MSGG_Left} \qquad
%\includegraphics[width=2.6in]{MSGG_Right}
%\caption{The local susceptibility functions corresponding to the left and right order parameter profiles (see figure \ref{fig:GGOP}) coexisting at capillary condensation at $\bar{x}_{h}=-236.2$ and $x_{t}=-25.795$.}
%\label{fig:MSGGOP}
%\end{figure}
%A pair of order parameter profiles coexisting at capillary condensation for $\bar{x}_{h}=-250$ and $x_{t}=-33.8105$ are presented in figure \ref{fig:RE}.

\section{Analytical results for the scaling behavior of the susceptibility}

\subsection{Exact results for the local susceptibility profiles in the case of nonzero field}

With respect to local susceptibility
\begin{equation}\label{nscafunchilocmt}
\chi_l(z|x_t,x_h)=L^{\gamma/\nu}X_\chi(\zeta|x_t,x_h),
\end{equation}
with $\nu=1/2$, $\gamma=1$ in the model under consideration, from Eqs. (\ref{localsusdefmt}) and (\ref{nscafunchilocmt}) one derives
\begin{equation}
X_\chi(\zeta|x_t,\bar{x}_h)=2\frac{\partial}{\partial \bar{x}_h}X_m(\zeta|x_t,\bar{x}_h).
\label{scalingfunctionsusceptibilitylocal}
\end{equation}
In particular, for the local susceptibility in the middle of the system one has 
\begin{equation}\label{hi12}
X_\chi(1/2|x_t,\bar{x}_h)=2 \dot{X}_{m0} (\bar{x}_h),
\end{equation}
where the dot indicates differentiation with respect to the variable $\bar{x}_h $.
According to Eq. (\ref{eqcontsrchimt}), $X_\chi$ satisfies the equation
\begin{equation}
X''_\chi(\zeta)-\left[x_t+6X_m^2(\zeta)\right]X_\chi(\zeta)=-1.
\label{eqXchi}
\end{equation}
On the other hand, differentiating the first integral (\ref{FirstIntA}) of the order parameter equation with respect to $\bar{x}_{h}$ and taking into account Eqs. (\ref{msaclingEqA}) and (\ref{IntConst}), 
% and assuming that $X_{m0}=X_{m0}(\bar{x}_{h})$ 
one obtains
\begin{equation} \label{MSEq20}
X'_{\chi }(\zeta )-\frac{X''_{m}(\zeta )}{X'_{m}(\zeta )} X_{\chi}(\zeta )+\frac{X_{m}(\zeta )-A}{X'_{m}(\zeta )}=0, \\
\end{equation}
where
\begin{eqnarray} \label{A}
A & = & \bar{x}_{h}\dot{X}_{m0}(\bar{x}_{h})+X_{m0}(\bar{x}_{h})\left[1-2\dot{X}_{m0}(\bar{x}_{h})\left(2X_{m0}^{2}(\bar{x}_{h})+x_t\right)\right] \nonumber \\
&=& X_{m0}(\bar{x}_{h}) -2\dot{X}_{m0}(\bar{x}_{h}) X_m''(1/2|x_t,\bar{x}_h).
\end{eqnarray}
The derivation of the solution of the linear first-order ordinary differential equation (\ref{MSEq20}) which meets the condition $ X_\chi\left( 0| x_t,\bar{x}_{h},X_{m0}\right)=0 $ is straightforward and so we can express the local susceptibility $X_{\chi }(\zeta ) $ through the order parameter $X_{m}(\zeta ) $ in the following explicit form
\begin{equation}\label{GS}
X_{\chi }(\zeta )=X_{m}^{\prime} (\zeta ) \int_{0}^{\zeta} \frac{A-X_{m}(w ) }{\left[X_{m}^{\prime }(w )\right]^2} dw.
\end{equation}
Starting from Eq. (\ref{scalingfunctionsusceptibilitylocal}) one can also determine the function $ \dot{X}_{m0}(\bar{x}_{h}) $. Obviously, Eq. (\ref{scalingfunctionsusceptibilitylocal}) can be written in the form 
\begin{equation} \label{NewExpLS}
\fl X_\chi\left( \zeta| x_t,\bar{x}_{h},X_{m0}\right) =2\frac{\partial}{\partial \bar{x}_h}X_m\left( \zeta| x_t,\bar{x}_{h},X_{m0}\right)
+2 \dot{X}_{m0}(\bar{x}_{h}) \frac{\partial}{\partial X_{m0}}X_m\left( \zeta| x_t,\bar{x}_{h},X_{m0}\right)
\end{equation}
and since $ X_\chi\left( 0| x_t,\bar{x}_{h},X_{m0}\right)=0 $ one derives  
\begin{equation}\label{NewExpLSConst}
\dot{X}_{m0}(\bar{x}_{h})=-\frac{\frac{\partial}{\partial \bar{x}_h}X_m\left( 0| x_t,\bar{x}_{h},X_{m0}\right)}{\frac{\partial}{\partial X_{m0}}X_m\left( 0| x_t,\bar{x}_{h},X_{m0}\right)}\cdot
\end{equation}
Thus, all the terms in Eq. (\ref{GS}) are completely determined only by means of the scaling function of the order parameter profile $X_m$ and its derivatives. Furthermore, Eq. (\ref{NewExpLS}) delivers an alternative analytical expression for $X_\chi$ provided one knows $X_m$. 

\begin{figure}[htbp]
\centering
\includegraphics[width=2.6in]{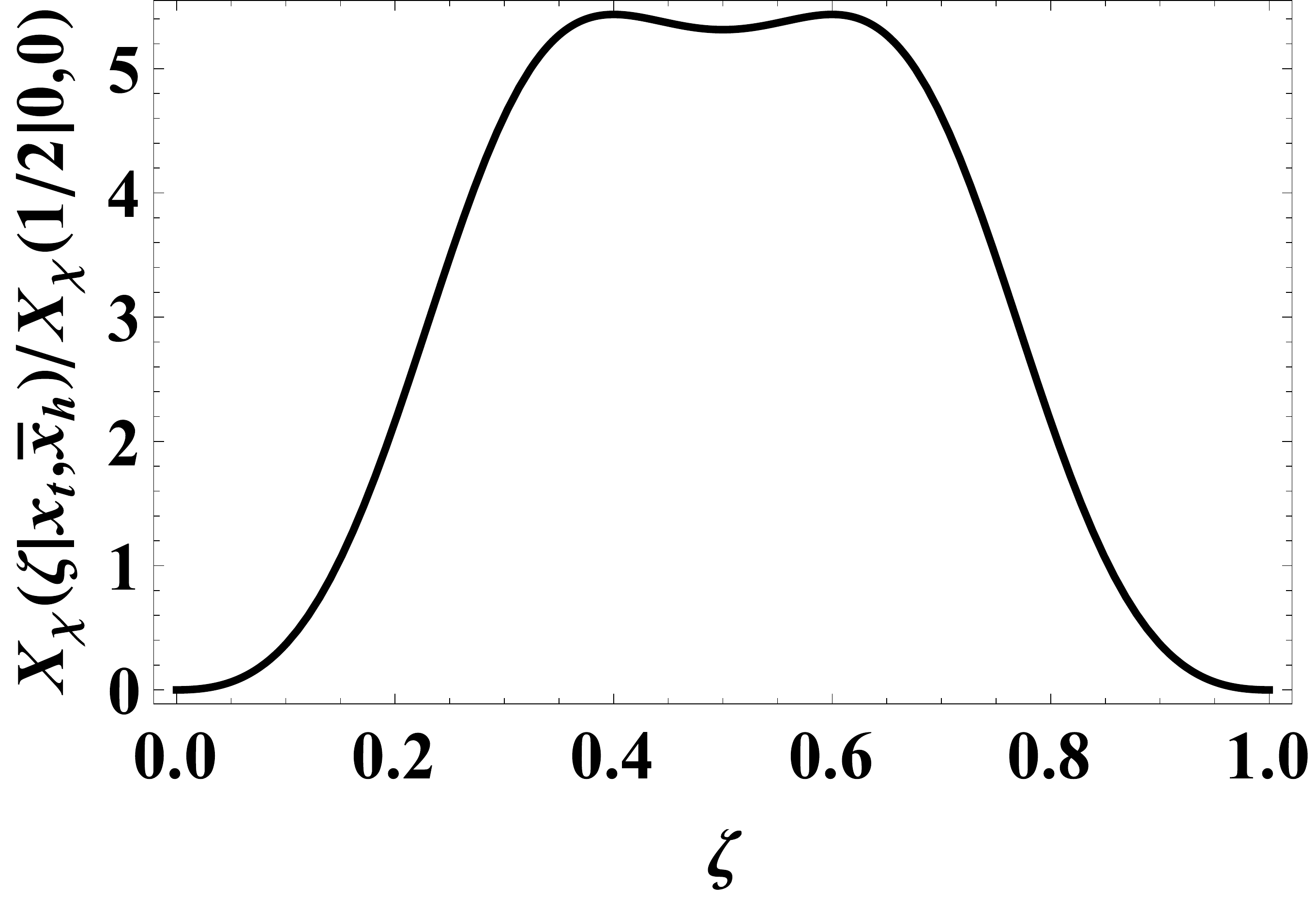} \qquad
\includegraphics[width=2.6in]{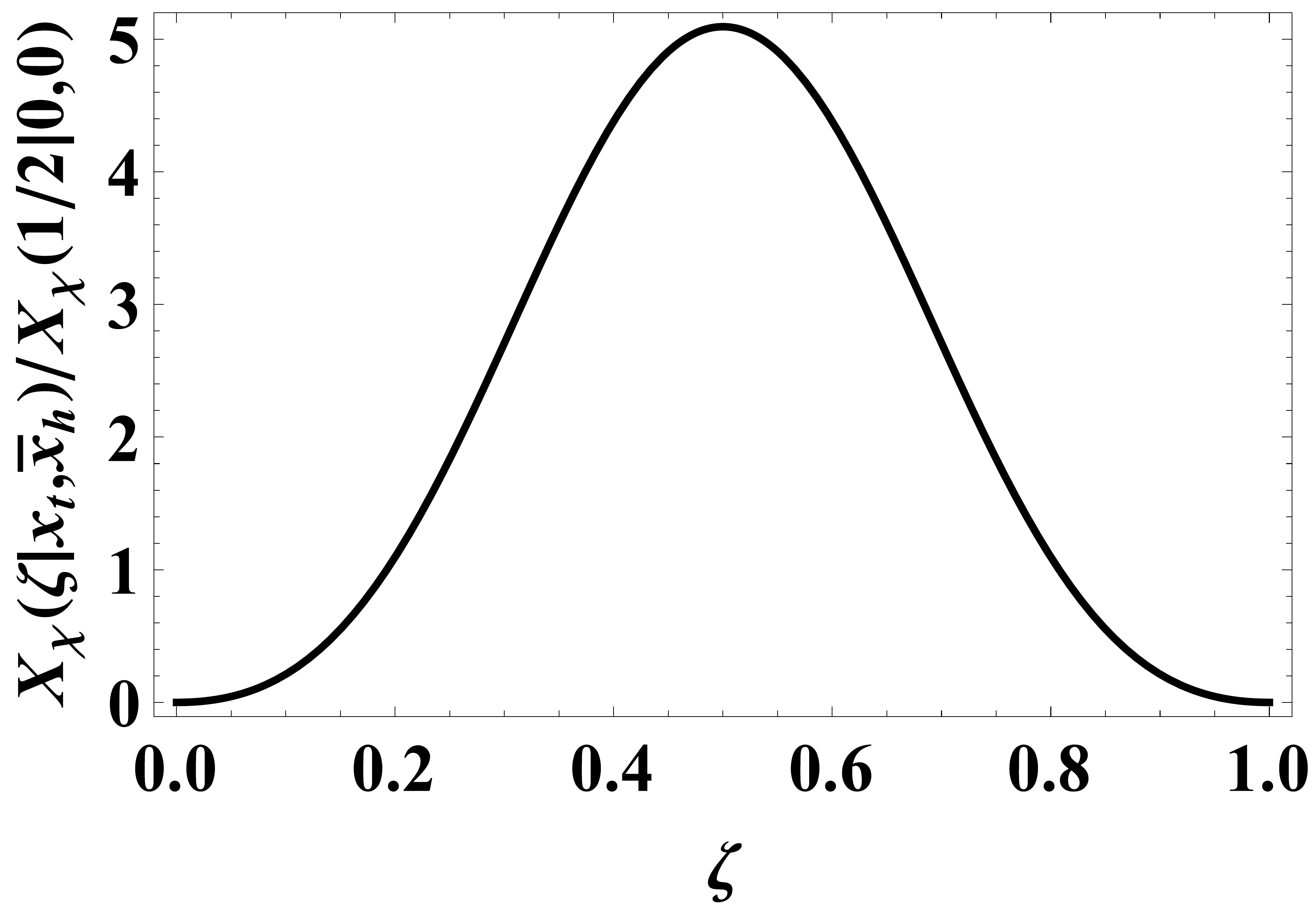}
\caption{The local susceptibility functions corresponding to the left and right order parameter profiles (see figure \ref{fig:RE}) coexisting at capillary condensation for $\bar{x}_{h}=-250$ and $x_{t}=-33.8105$.} 
\label{fig:MSEx1}
\end{figure}

The local susceptibility functions corresponding to the pair of order parameter profiles depicted in figure \ref{fig:RE} that are coexisting at capillary condensation curve  are obtained using Eqs. (\ref{NewExpLS}) and (\ref{NewExpLSConst}) and are presented in figure \ref{fig:MSEx1}.
From this figure we see that on the capillary condensation curve the local susceptibility might have either one ot two local maxima. The inspection of figures \ref{fig:RE} and \ref{fig:MSEx1} let us to conclude that in the case when the local susceptibility is characterized by two symmetrical local maxima they are centered, approximately, around the two gas-liquid interfaces in the system.
%The values of the total susceptibility normalized with $ X_\chi\left( 1/2| 0,0\right) $, corresponding to the local susceptibility profiles presented in figure \ref{fig:MSEx1}, are $6.095$ in the ''left'' case, and $4.378$ in the ''right'' one. 

\subsection{Exact results for the local susceptibility profiles in the case of zero field}

When $h=0$ one can determine the scaling function  $X_\chi(z|x_t)\equiv X_\chi(z|x_t,x_h=0)$ (see Eq. (\ref{nscafunchilocmt})) of the local susceptibility 
in an explicit analytical form  \cite{DRB2009}. One has
\begin{equation}\label{Xchigenmt}
X_\chi(\zeta|x_t)=\psi_i (\zeta|x_t)+c_2 \psi_2 (\zeta|x_t),
\end{equation}
where
\begin{equation}\label{partsolexplmt}
\psi_i(\zeta|x_t)=-\frac{k'^{\,2}}{X_{m,0}^2}\left\{1-2\,{\rm dn}\left[i \frac{X_{m,0}}{k'}\,\zeta;k'\right]^2\right\},
\end{equation}
and
\begin{eqnarray}
\label{secondsolexplmt}
&& \psi_2(\zeta|x_t)  = -\frac{k'}{k^2 X_{m,0}^3}\left\{ {\rm dn}\left(i \frac{X_{m,0}}{k'}\,\zeta;k' \right) {\rm sn}\left(i \frac{X_{m,0}}{k'}\,\zeta;k' \right) \right.
\\
&&\left. \times \left[k'(1-2k'^2){\rm E}\left({\rm am}\left(i \frac{X_{m,0}}{k'}\,\zeta;k' \right);k'\right)-ik^2X_{m,0}\zeta\right] \right.\nonumber  \\
&& \left.\qquad \quad +k'{\rm cn}\left(i \frac{X_{m,0}}{k'}\,\zeta;k' \right) \left[k'^2+(1-2k'^2)\;{\rm dn}\left(i \frac{X_{m,0}}{k'}\,\zeta;k' \right)2\right]\right\} \nonumber, 
\end{eqnarray}
with
\begin{equation}\label{c2explmt}
c_2(x_t)=\frac{4 k'^2 k^2 K\left(k\right)}{k'^2
   K\left(k\right)+\left(k^2-k'^2\right) E\left(k\right)}.
\end{equation}
Here $E(k)$ is the complete elliptic integral of the second kind.
\begin{figure}[htbp]
\centering
\includegraphics[width=2.8in]{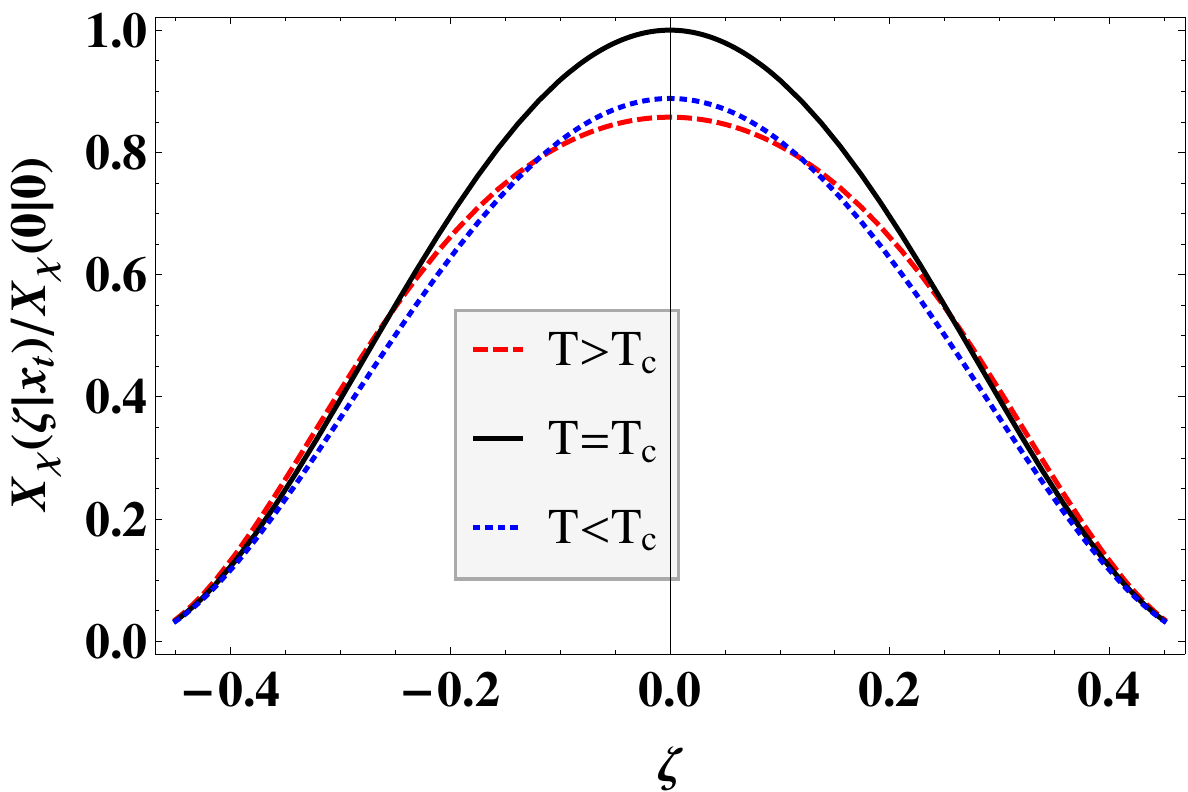}
\caption{Color online. Plot of the function $X_\chi(\zeta|x_t)$ normalized per value of this function in the middle of the system for $T=T_c$.}\label{fig:XchiTc}
\end{figure}

The behavior of the scaling function $X_\chi$ for three different temperatures -- well below, at and well above $T_c$ is shown on figure \ref{fig:XchiTc}.

\subsection{Exact results for the total susceptibility in the case of nonzero field}

According to Eq. (\ref{chifss}), for the scaling function of the total susceptibility $ X\left(x_t,x_h\right) $ one has 
\begin{equation}\label{schifss}
X\left(x_t,x_h\right)=\int_{0}^{1} X_\chi \left(\zeta|x_t,x_h\right) d\zeta,
\end{equation}
wherefrom one obtains that
\begin{equation}\label{schifss_res}
X\left(x_t,x_h\right)=2\int_0^{1/2} \frac{ [X_{m} (\zeta )-X_{m0}] [X_{m}(\zeta )-A] }{\left[X_{m}^{\prime }(\zeta )\right]^2} d\zeta.
\end{equation}

\begin{figure}[htbp]
\centering
\includegraphics[width=2.9in]{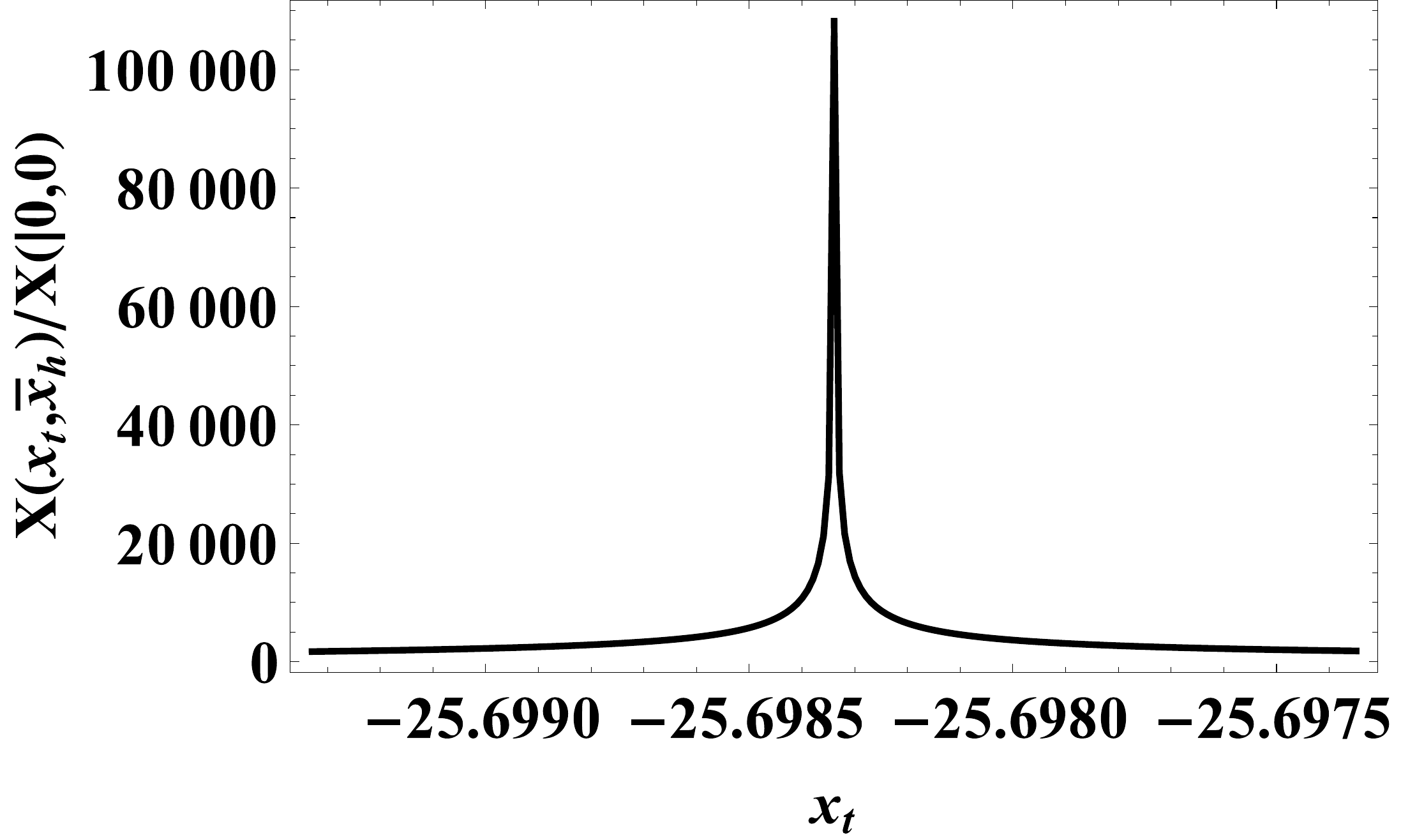}\quad
\includegraphics[width=2.6in]{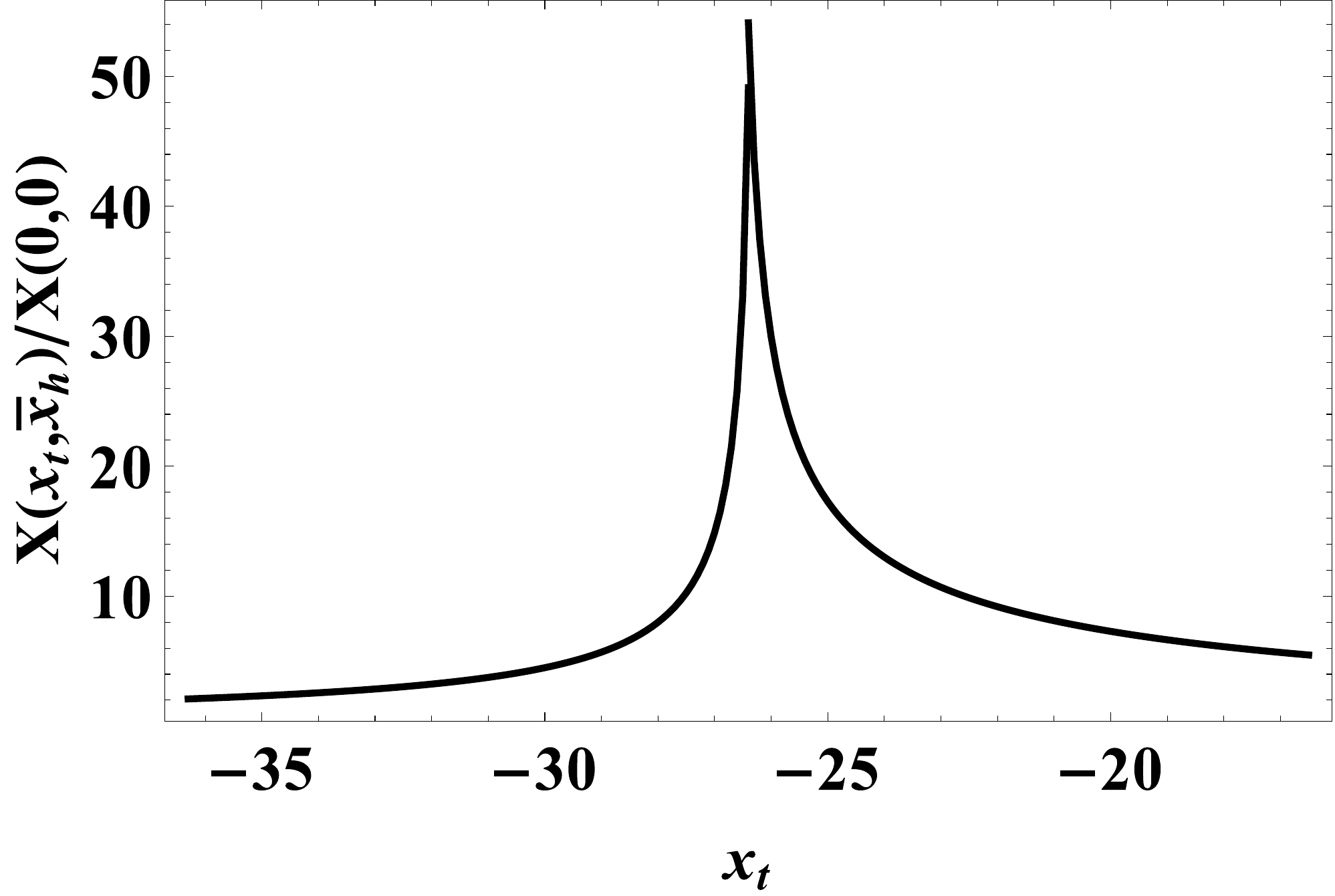}
\caption{The scaling functions of the total susceptibility $X(x_t,-236.0350005)$, i.e., near $T_{c,L}$ (left), and $X(x_t,-237.24)$, i.e., near $T_{\rm cap}$ (right), normalized with the total susceptibility$X(0,0)$ at the bulk critical point. $ X\left(x_t,x_h\right)$ shows a clear singularity at $T_{c,L}$. The small finite jump of $X\left(x_t,x_h\right)$ near $T_{\rm cap}$ happens when $x_t$   passes through the pre-capillary condensation line.}
\label{fig:TS_GG_GT}
\end{figure}

\begin{figure}[htbp]
\centering
\includegraphics[width=2.5in]{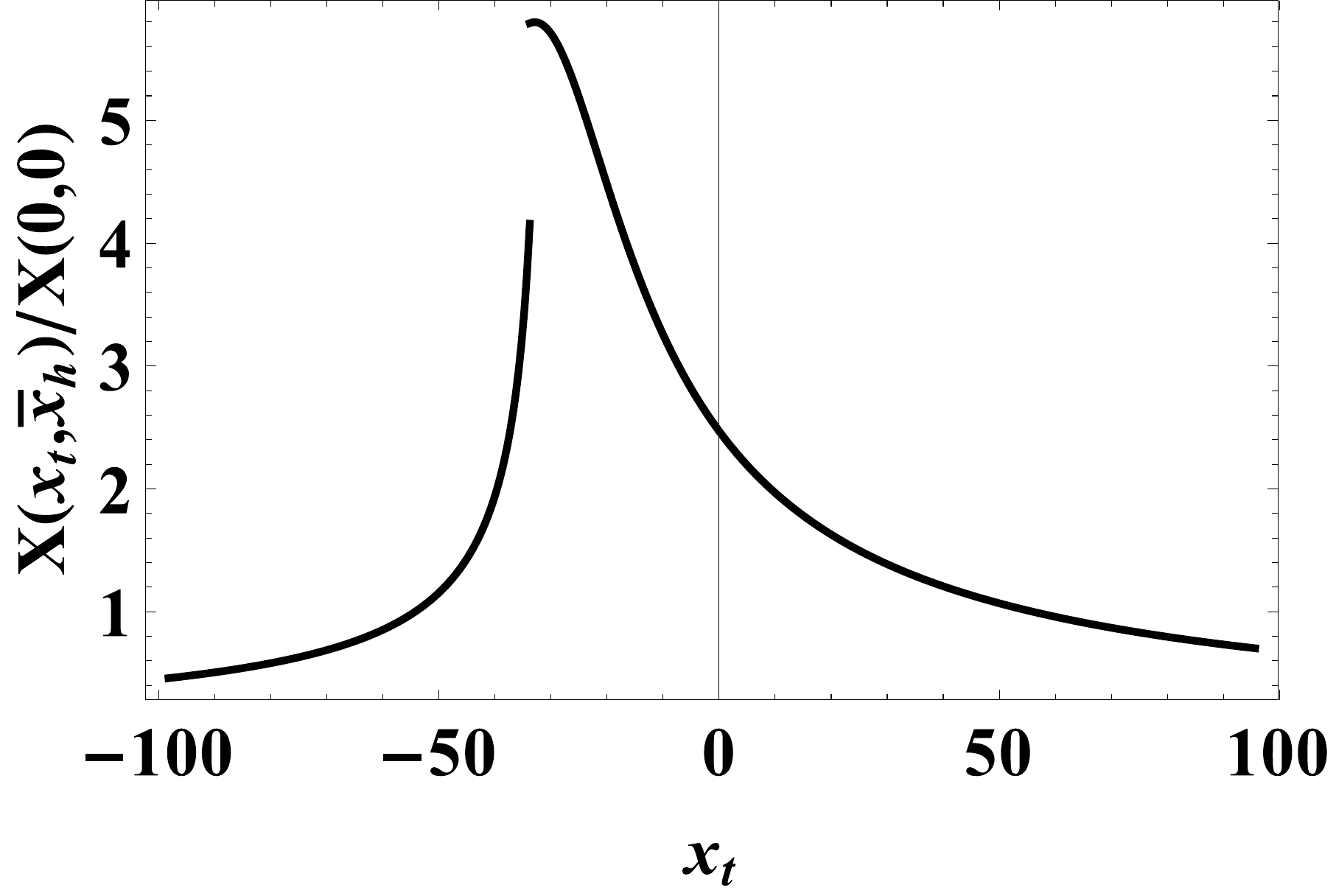}\qquad
\includegraphics[width=2.6in]{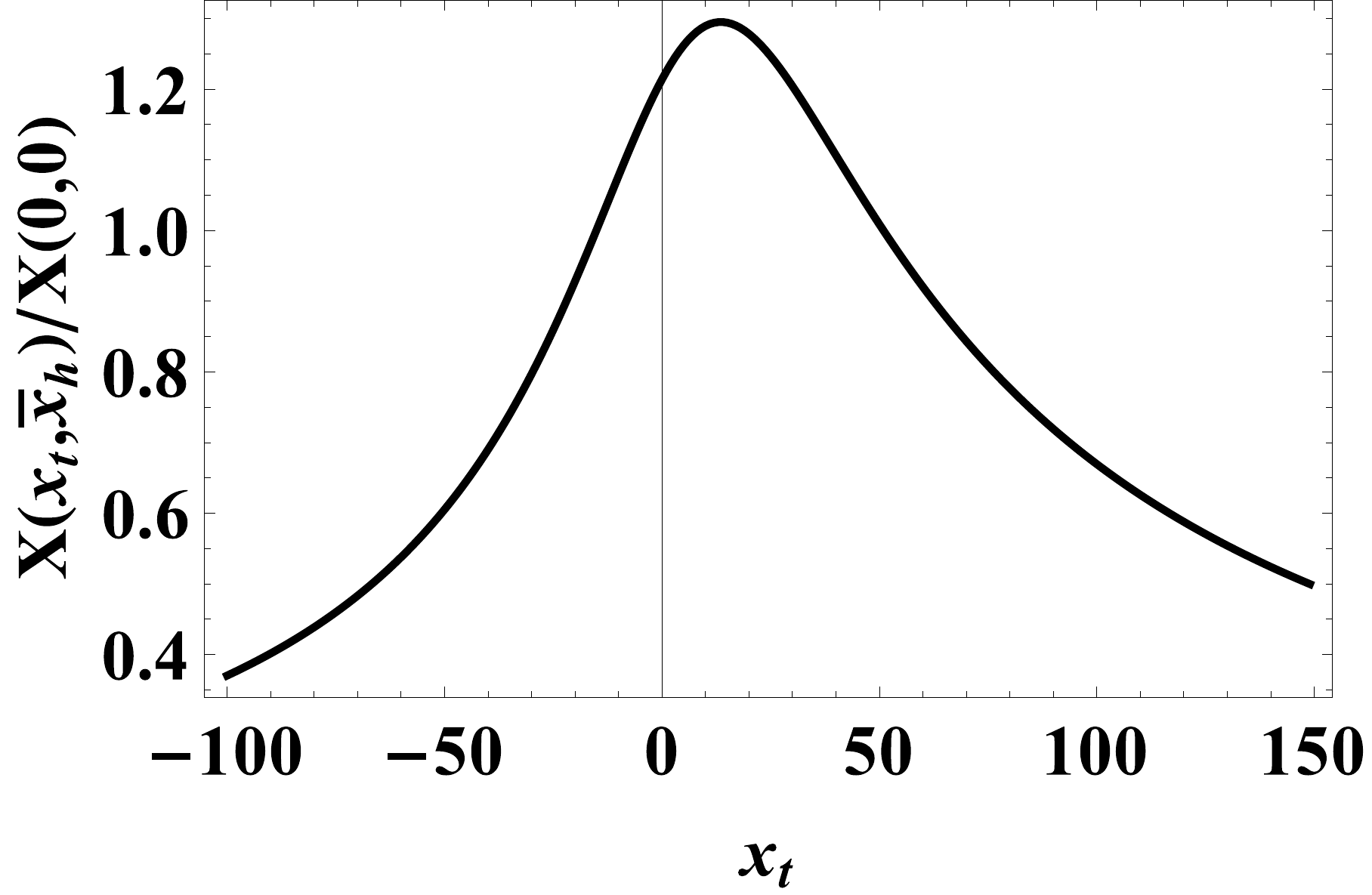}
\caption{The typical temperature behavior of the scaling function of the total susceptibility  $X(x_t,-250)$  (left) and $X(x_t,-50)$ (right) well below and well above  the critical value of the magnetic scaling variable $ x_h^{(c)}$, correspondingly, normalized with the total susceptibility$X(0,0)$ at the bulk critical point. }
\label{TSh-250-50}
\end{figure}

The temperature behavior of the susceptibility for several fixed values of the field scaling variable are given in figures  \ref{fig:TS_GG_GT} and \ref{TSh-250-50}. 
Figure \ref{fig:TS_GG_GT} illustrates the dependence of the total susceptibility in the vicinity of the points $T_{c,L}$ and $T_{\rm cap}$.  The left sub-figure there shows the variation of the total susceptibility in the vicinity of the critical point $T_{c,L}\equiv (x_t^{(c)},x_h^{(c)})=(-25.6983391, -236.0350005)$ of the finite system, while the right one shows its behavior around the capillary condensation point $T_{\rm cap}\equiv (x_t^{\rm(cap)},x_h^{\rm(cap)})=(-26.4025, -237.2395)$. Here $ x_t^{\rm(cap)} $ is understood as  the highest temperature at which the entire capillary fills with liquid.  Figure \ref{TSh-250-50} 
represents the variation of the total susceptibility with the temperature for fields far away from $ x_h^{(c)}$. The left figure corresponds to $\bar{x}_h = -250 < x_h^{(c)}$ and the finite jump of $ X\left(x_t,x_h\right) $ indicates the passing of $x_t$ through the capillary condensation line. The curve on the right figure demonstrates that $ X\left(x_t,x_h\right)$, as expected, is smooth when the field $\bar{x}_h = -50$ is well above $ x_h^{(c)}$.

\subsection{Exact results for the total susceptibility in the case of zero field}

Having in mind Eqs. (\ref{Xchigenmt}) -- (\ref{c2explmt}), one can also determine \cite{DRB2009} the scaling function $X(x_t)\equiv X(x_t,0)$ of the total susceptibility $\chi(x_t)$, where 
\begin{equation}\label{scafunchimt}
\chi(x_t)=L^{\gamma/\nu} X(x_t),
\end{equation}
with $\nu=1/2$ and $\gamma=1$ for the model considered. The corresponding result for $X(x_t)$ is
\begin{equation}\label{Xfinalmt}
X(x_t)=\frac{c_2(x_t)/K\left(k\right)+K\left(k\right)-2 E\left(k\right)}{4
   K^3\left(k\right)}.
\end{equation}
Here $c_2$ is given in Eq. (\ref{c2explmt}) and $x_t$ is to be determined from Eqs. (\ref{tk}) and (\ref{tkb}). 
\begin{figure}[htbp]
\centering 
\includegraphics[width=2.8in]{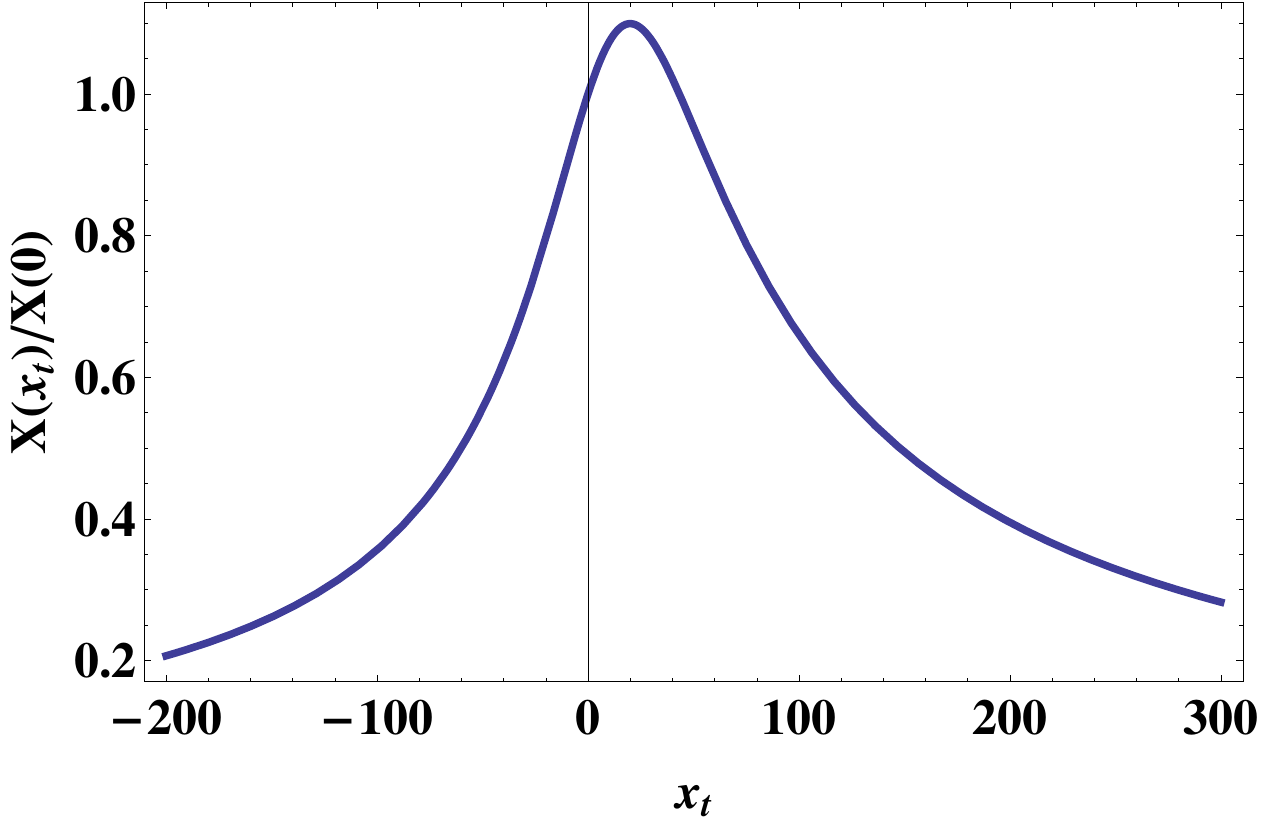}
\caption{The scaling function of the total susceptibility $X(x_t)$ normalized per its value at the bulk critical point.}
\label{fig:totsuscont}
\end{figure}

The behavior of $X(x_t)$ is illustrated in figure  \ref{fig:totsuscont}. One observes that $X$ possesses a maximum above the critical temperature $T_c$ of the bulk system at $x_t=19.9678$ and the value of the maximum is $1.09925$ times higher than the value of the total susceptibility at the critical point. 

\section{Discussion and concluding remarks}

In the current article we present exact {\it analytical} results for the temperature-field behavior  of the order parameter profile and for the behavior of the main response functions -- local and total susceptibilities, for one of the basic and most studied models in the statistical mechanics -- the mean-field Ginsburg-Landau $\phi^4$ model. We have studied the properties of this model under the so-called $(+,+)$ boundary conditions for a system with a film geometry in the case when both bounding the system surfaces strongly prefer the liquid phase of the confined fluid system. We studied both the critical regime $tL^{1/\nu}={\cal O}(1)$, $hL^{\Delta/\nu}={\cal O}(1)$, where $t=(T-T_c)/T_c$, $\nu=1/2$, $\Delta/\nu=3$, and the capillary condensation regime $T<T_c, h<0$. The basic new exact result is the one derived for the behavior of the scaling function of the order parameter profile presented in Eq. (\ref{WSA}), wherefrom one derives expressions (\ref{A}) -- (\ref{NewExpLSConst}) for the scaling function of the local susceptibility, and \eq{schifss_res} for the total susceptibility. Analytical results for the behavior of these quantities were known before only for the $h=0$ case. The behavior of the order parameter profile for different values of $T$ and $h$ is visualized in figures  \ref{fig:Xm}, \ref{fig:Ex1}, \ref{fig:GGOP} and \ref{fig:RE}, of the local susceptibility -- in figures \ref{fig:MSEx1} and \ref{fig:XchiTc}, and that one of the total susceptibility - in figures \ref{fig:TS_GG_GT}, \ref{TSh-250-50} and \ref{fig:totsuscont}.

Based on the derived exact analytical expressions we obtained the phase diagram (see figure \ref{fig:DP}, left) and the coordinates of the critical point $T_{c,L}\equiv (x_t^{(c)},x_h^{(c)})=(-25.6983391, -236.0350005)$ which are in excellent agreement with those determined in \cite{SHD2003}, see Fig 13 therein \cite{rem}. We observed that along the coexistence line between $T_{c,L}$ and the the capillary condensation point $T_{\rm cap}\equiv (x_t^{\rm(cap)},x_h^{\rm(cap)})=(-26.4025, -237.2395)$ the system jumps from a less dense gas to a more dense gas (see figure \ref{fig:GGOP})
before switching on continuously into the usual jump from gas to liquid state in the middle of the system in the capillary condensation regime -- see figure \ref{fig:RE}. This is an unexpected theoretical result that calls for an experimental check-up --- one shall see if it is a theoretical artifact of the considered model, or corresponds to experimentally observable phenomena. 

Closing this discussion, let us also mention that the mean-field solutions are exact for the critical behavior of systems with dimensionality $d\ge 4$ (apart from some logarithmic corrections for $d=4$). Next, these solutions  serve as a starting point for more sophisticated analytical techniques
like the renormalization group calculations utilizing the $\varepsilon$-expansion \cite{D86,K97,P90}. Thus, our results shall be also helpful for such future theoretical considerations. 

\section*{Acknowledgements}

We wish to thank J. Rudnick, S. Dietrich, R. Evans and A. Maci{\`o}{\l}ek for helpful comments and the critical reading of the manuscript of this article. 

\appendix

\section*{Appendix. Some details on the numerical evaluations}
%\label{appendix}

\setcounter{section}{1}

In the current appendix we present a numerical verification of \eq{eq:energy_fiff} and provide some technical details related to the determination of the numerical results presented in some of the plots in the current article. 

\begin{figure}[htbp]
\centering
\includegraphics[width=2.86in]{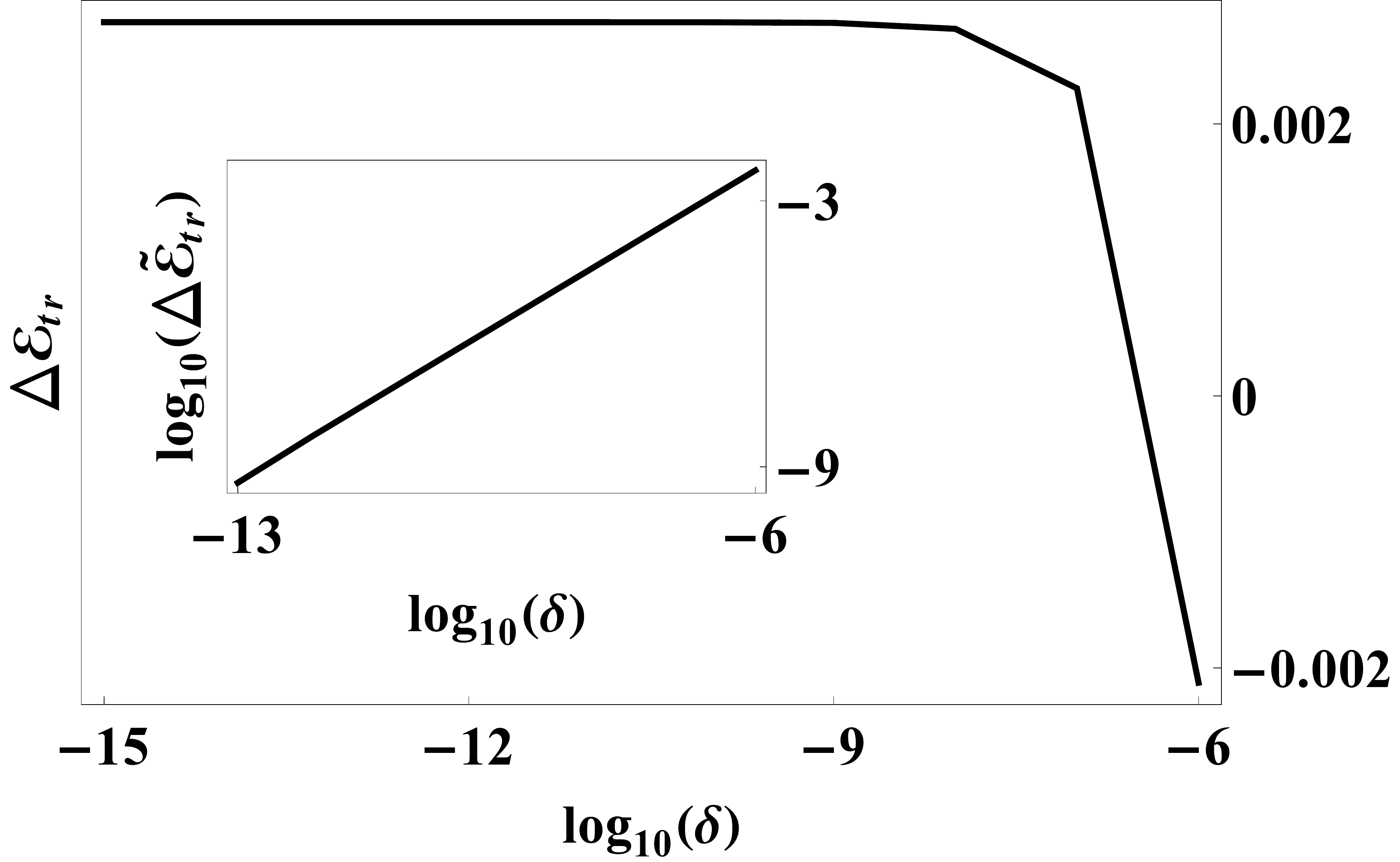}\qquad
\includegraphics[width=2.9in]{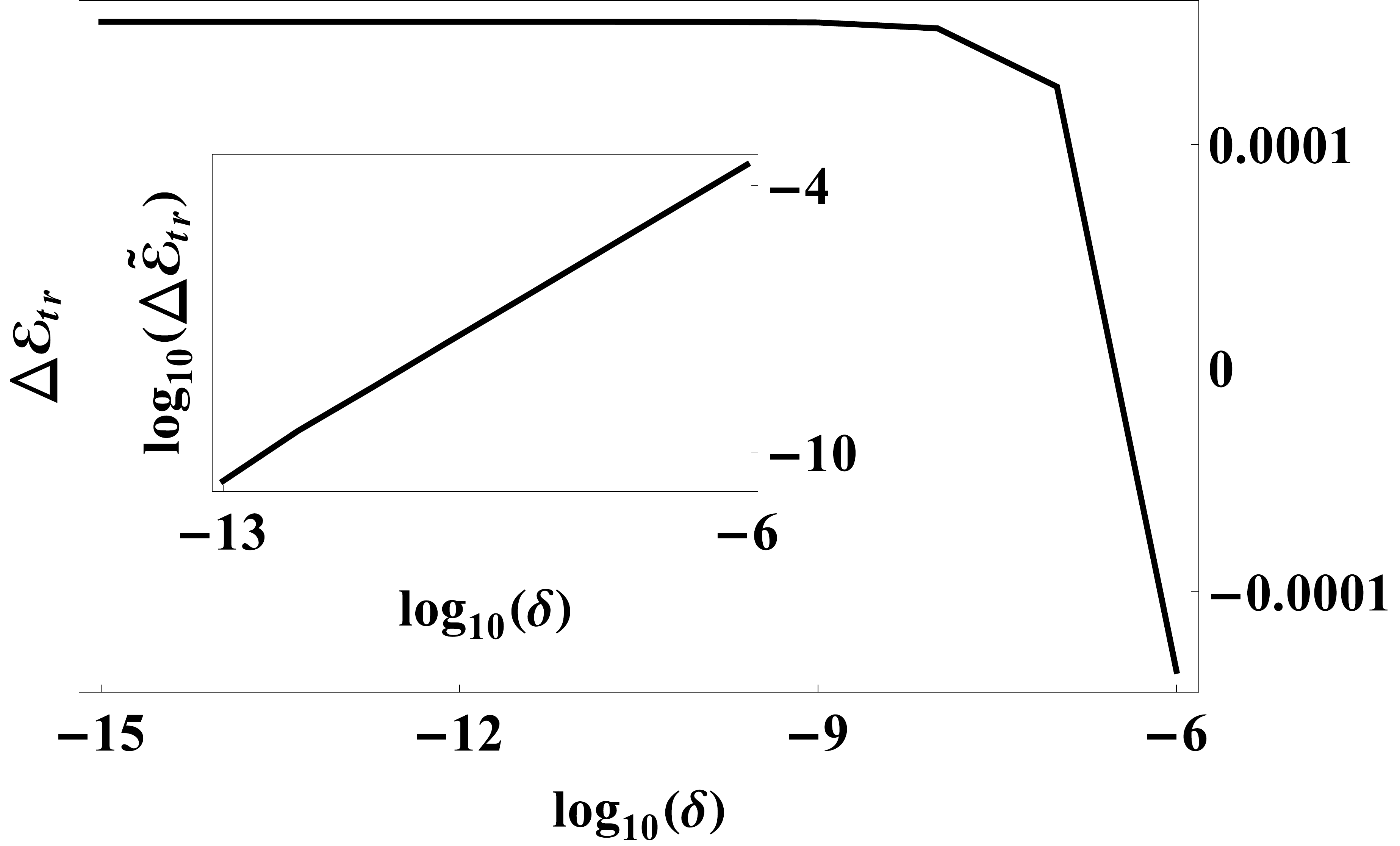}
\caption{The evolution of the difference $\Delta \mathcal{E}_{tr}$ between the truncated energies of the two competing states for $10^{-15} \leq \delta \leq 10^{-6}$ at $x_t = -80,\bar{x}_h = -331.89$ (left) and $x_t = -26,\bar{x}_h = -236.55$ (right). }
\label{fig:DeltaEvolution}
\end{figure}

As already stated in the main text, using \eq{eq:energy_fiff} the difference $\Delta \mathcal{E}$ between the energies of two states is determined by the difference between the two respective truncated energies up to terms of order ${O}(\delta) $, i.e. 
\begin{equation}
\label{eq:deltaE_A}
\Delta \mathcal{E}=\Delta \mathcal{E}_{tr}(\delta) +{O}(\delta), 
\end{equation}
where 
\begin{eqnarray}
\Delta \mathcal{E} & = & \mathcal{E}(x_{t},\bar{x}_{h},X_{m02})-\,\mathcal{E}(x_{t},\bar{x}_{h},X_{m01}), \\
\Delta \mathcal{E}_{tr}(\delta)  & = & \mathcal{E}_{tr}(x_{t},\bar{x}_{h},X_{m02};\delta)-\,\mathcal{E}_{tr}(x_{t},\bar{x}_{h},X_{m01};\delta).
\end{eqnarray}
The relation (\ref{eq:deltaE_A}) means that {i)} $\delta$ determines the precision with which  we determine the energy difference $\Delta \mathcal{E}$  between any two solutions of the oder parameter problem and {ii)} that $\lim_{\delta \to 0} \Delta \mathcal{E}_{tr}(\delta)=\Delta \mathcal{E}$.  A numerical verification of \eq{eq:energy_fiff} is presented in figure \ref{fig:DeltaEvolution}.  
This figure comprises two typical examples of the evolution of $\Delta\mathcal{E}_{tr}$ as a function of $\delta$  for $x_t = -80,\bar{x}_h = -331.89$ (left) and $x_t = -26,\bar{x}_h = -236.55$ (right).
% for the two competing there states of the system characterized by two different values of $X_{m0}$.
%Please note that the choice of $\delta$ is indeed important. Choosing $\delta$ too large can lead to wrong conclusions about the energy relation between the competing states.
The choice of $\delta$ is indeed important because choosing $\delta$ too large can lead to a wrong conclusion about the energy difference of the competing states.
As figure \ref{fig:DeltaEvolution} shows, the energy difference is negative for $\delta = 10^{-6}$, but becomes positive at $\delta = 10^{-7}$ and stabilizes below $\delta = 10^{-8}$. Thus, one should carefully choose the truncation $\delta$ in equations (\ref{TrunkatedEnergy}) and (\ref{eq:energy_fiff}) in order to ensure that the difference of the truncated energies properly indicates which of the two competing states for a given $(x_t,\bar{x}_h)$ combination is of less energy. To find such $\delta$ one, e.g., chooses $\delta = 10^{-m}$ and computes the difference of the truncated energies for increasing integers $m$ till this energy no longer changes sign. Figure \ref{fig:DeltaEvolution}  depicts how  the value of $\Delta \mathcal{E}_{tr}$ approaches its limiting value $\Delta \mathcal{E}$ with the decrease of $\delta$, in accordance with Eqs. (\ref{eq:energy_fiff}) and (\ref{eq:deltaE_A}). The insets therein show the variation of the difference 
 $\Delta\tilde{\mathcal{E}}_{tr}=\log_{10}\left(\Delta\mathcal{E}_{tr}(10^{-14})-\Delta \mathcal{E}_{tr}(\delta)\right)$ with $\log_{10}(\delta)$.
As we see, this dependence is linear which is exactly what it shall be expected on the basis of \eq{eq:energy_fiff}. 
%of  On the vertical axis the difference $\mathcal{E}_{tr}\equiv \Delta \mathcal{E}_{tr}-\Delta \mathcal{E}$ is plotted, where $\Delta \mathcal{E}$ is considered to coincide, with a precision of $10^{-14}$, with the corresponding value of $\Delta \mathcal{E}_{tr}$ for $\delta = 10^{-14}$.  As we see, the dependence of $\log_{10} (\mathcal{E}_{tr})$ on $\log_{10}(\delta)$ is linear. The last is exactly what it shall be expected on the basis of \eq{eq:energy_fiff}. 

\begin{figure}[htbp]
\centering
\includegraphics[width=2.8in]{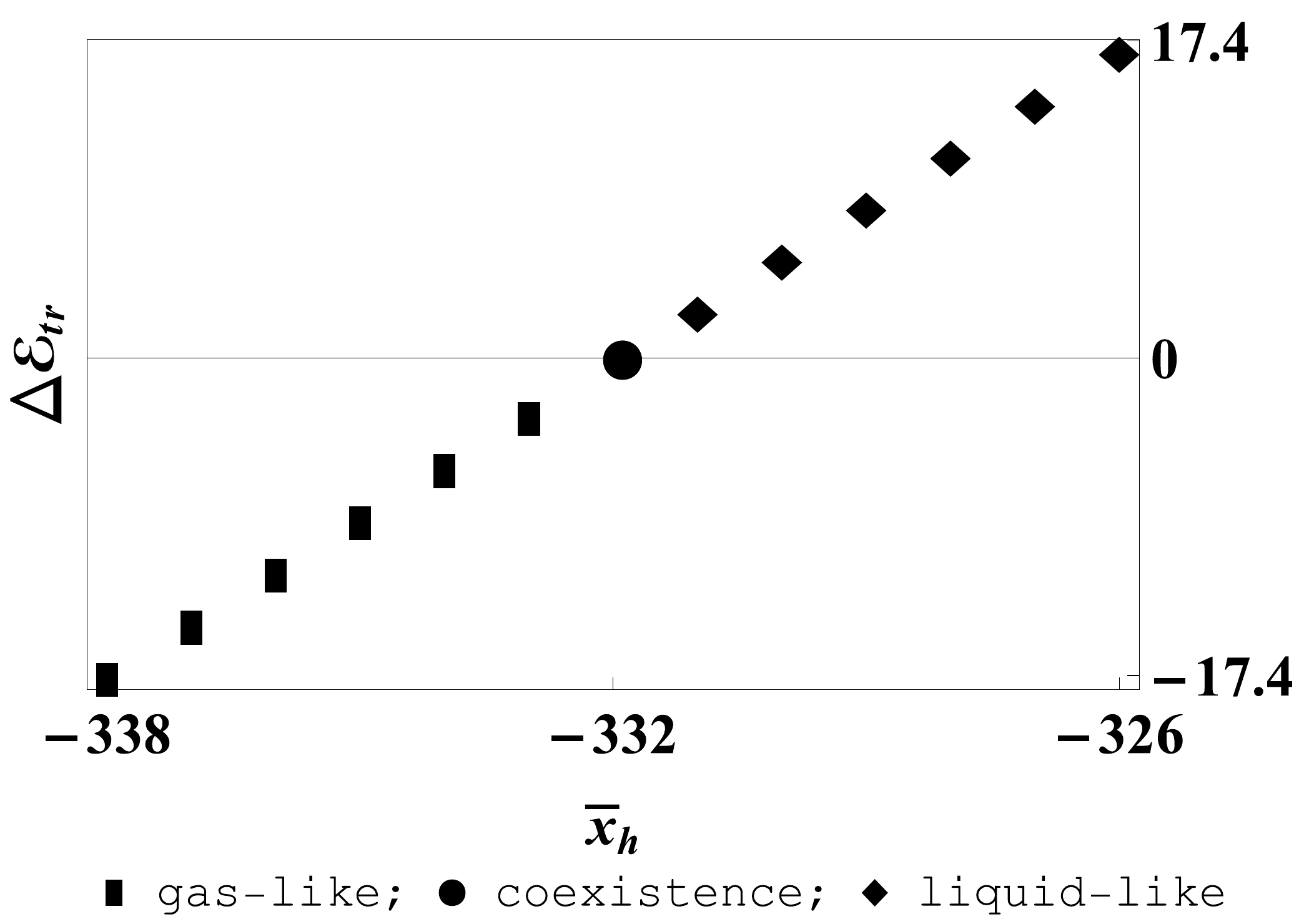}\qquad
\includegraphics[width=2.8in]{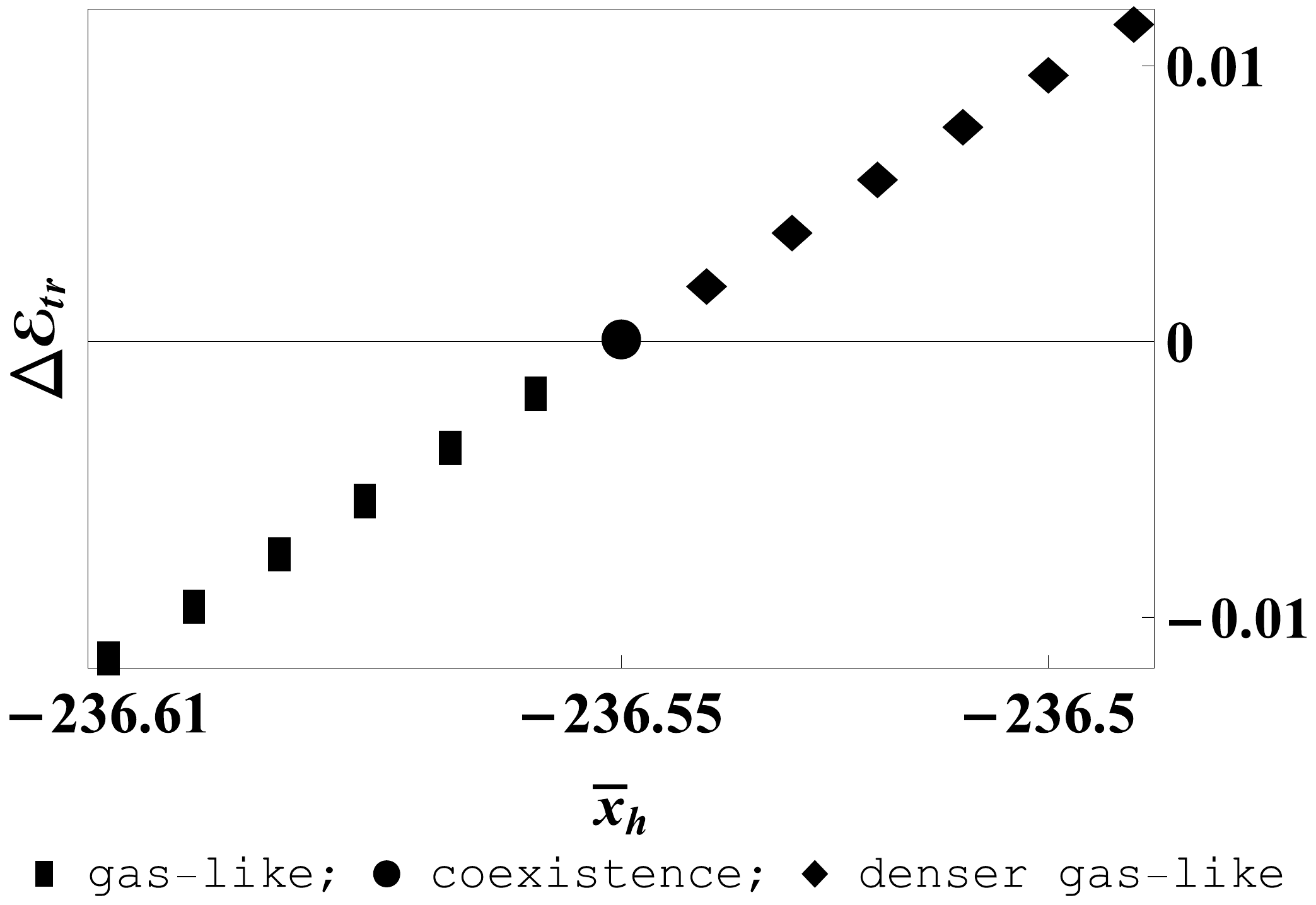}
\caption{The evolution of the difference $\Delta\mathcal{E}_{tr}$ between the energies of the two competing states for $x_t = -80$ (left; the usual jump from gas-like state to liquid-like state), and $x_t = -26$ (right; a jump from a gas-like state to a denser gas-like state). The circles indicate the coexistence points for these temperatures.}
\label{fig:EEvolution}
\end{figure}

At the end, let us present some clarification remarks about the procedure followed in the determination of the phase diagram given in figure \ref{fig:DP}. If, at any fixed $(x_t,\bar{x}_h)$, the system possesses two competing states we determine the one which is stable by choosing that one with less energy. An illustration of the evolution of $\Delta \mathcal{E}_{tr}$ with the change of the thermodynamic parameters governing the behavior of the system is shown in figure \ref{fig:EEvolution}. 
The point $(x_t,\bar{x}_h)$ for which $\Delta \mathcal{E}_{tr}$ vanishes within the chosen precision $\delta$, is the point that belongs to the phase separation line of the phase diagram. The last implies that the phase diagram is, of course, also determined within that precision. In the current article, as stated above, we have worked with $\delta = 10^{-14}$.

%\newpage 

%\section*{References}

%\bibliography{publications}
%\bibliographystyle{iopart-num}

\providecommand{\newblock}{}

\end{document}